\documentclass[a4paper,fleqn,usenatbib]{mnras}

\usepackage[T1]{fontenc}
\usepackage{ae,aecompl}
\usepackage{graphicx}	
\usepackage{amsmath}	
\usepackage{amssymb}	

\newcommand{\oiidoub}{[\textrm{O}\textsc{ii}]\ensuremath{\lambda3727,3729}}

\newcommand{\oiiidoub}{[\textrm{O}~\textsc{iii}]\ensuremath{\lambda\lambda4959,5007}}
 
\newcommand{\ha}{\ifmmode {\rm H}\alpha \else H$\alpha$\fi}
\newcommand{\hb}{\ifmmode {\rm H}\beta \else H$\beta$\fi}
\newcommand{\lya}{\ifmmode {\rm Ly}\alpha \else Ly$\alpha$\fi}
\newcommand{\pg}{\ifmmode {\rm P}\gamma \else Pa$\gamma$\fi}
\newcommand{\lyb}{\ifmmode {\rm Ly}\beta \else Ly$\beta$\fi}
\newcommand{\lyg}{\ifmmode {\rm Ly}\gamma \else Ly$\gamma$\fi}
\newcommand{\ciii}{\textrm{C}\textsc{iii}]\ensuremath{\lambda1908}}
\newcommand{\ciiidoub}{\textrm{C}\textsc{iii}]\ensuremath{\lambda\lambda1907,1909}}
\newcommand{\siii}{[\textrm{Si}\textsc{ii}]\ensuremath{\lambda1260}}
\newcommand{\oi}{[\textrm{O}\textsc{i}]\ensuremath{\lambda1303}}
\newcommand{\cii}{[\textrm{C}\textsc{ii}]\ensuremath{\lambda1334}}

\newcommand{\civ}{\textrm{C}\textsc{iv}\ensuremath{\lambda1548,1550}}

\newcommand{\heii}{\textrm{He}\textsc{ii}\ensuremath{\lambda1640}}
\newcommand{\oiiiuv}{\textrm{O}\textsc{iii}]\ensuremath{\lambda1661,1666}}

\def\kms{km s$^{-1}$}

\def\ergs{\ifmmode \mathrm{erg\hspace{1mm}s}^{-1} \else erg s$^{-1}$\fi}
\def\ergscm{erg s$^{-1}$ cm$^{-2}$}
\def\micron{\ifmmode \mu\mathrm{m} \else $\mu$m\fi}
\def\msun{\ifmmode \mathrm{M}_{\odot} \else M$_{\odot}$\fi}
\def\msunyr{\ifmmode \mathrm{M}_{\odot} \hspace{1mm}{\rm yr}^{-1} \else $\mathrm{M}_{\odot}$ yr$^{-1}$\fi}
\def\zsun{\ifmmode Z_{\odot} \else Z$_{\odot}$\fi}
\def\lsun{\ifmmode L_{\odot} \else L$_{\odot}$\fi}
\def\mstar{\ifmmode \mathrm{M}_{\star} \else M$_{\star}$\fi}

\title[Ly$\alpha$ nebulae]{Illuminating gas in-/outflows in the MUSE deepest fields: discovery of
Ly$\alpha$ nebulae around forming galaxies at $z\simeq3.3$}

\author[E.~Vanzella et al.]{
E.~Vanzella$^{1}$\thanks{E-mail: eros.vanzella@oabo.inaf.it},
I.~Balestra$^{2,3}$,
M.~Gronke$^{4}$, 
W.~Karman$^{5}$,
G.B.~Caminha$^{6}$, \newauthor 
~M.~Dijkstra$^{4}$, 
P.~Rosati$^{6}$, 
S.~De Barros$^{1,7}$,
K.~Caputi$^{5}$,
C.~Grillo$^{8}$, 
P.~Tozzi$^{9}$, \newauthor  
~M.~Meneghetti$^{1}$,
A.~Mercurio$^{10}$,
R.~Gilli$^{1}$\thanks{Based on observations collected at the European Southern Observatory for Astronomical research in the
Southern Hemisphere under ESO programmes P096.A-0045 and P094.A-0115.}
\\
\\
$^{1}$INAF -- Osservatorio Astronomico di Bologna, via Ranzani 1, 40127 Bologna, Italy\\
$^{2}$University Observatory Munich, Scheinerstrasse 1, 81679 Munich, Germany\\
$^{3}$INAF -- Osservatorio Astronomico di Trieste, via G. B. Tiepolo 11, I-34143, Trieste, Italy\\
$^{4}$Institute of Theoretical Astrophysics, University of Oslo, Postboks 1029 Blindern, NO-0315 Oslo, Norway\\
$^{5}$Kapteyn Astronomical Institute, University of Groningen, Postbus 800, 9700 AV Groningen, The Netherlands\\
$^{6}$Dipartimento di Fisica e Scienze della Terra, Universit\`a di Ferrara, via Saragat 1, 44122 Ferrara, Italy\\
$^{7}$Observatoire de Gen\`eve, Université de Gen\`eve, 51 Ch. des Maillettes, 1290, Versoix, Switzerland\\
$^{8}$Dark Cosmology Centre, Niels Bohr Institute, University of Copenhagen, Juliane Maries Vej 30, DK-2100 Copenhagen, Denmark\\
$^{9}$INAF -- Osservatorio Astrofisico di Arcetri, Largo E. Fermi, I-50125, Firenze, Italy\\
$^{10}$INAF -- Osservatorio Astronomico di Capodimonte, Via Moiariello 16, I-80131 Napoli, Italy\\
}

\date{}

\pubyear{2016}

\begin{document}
\label{firstpage}
\pagerange{\pageref{firstpage}--\pageref{lastpage}}
\maketitle

\begin{abstract}
We report on the discovery of extended Ly$\alpha$ nebulae at $z\simeq3.3$ 
in the Hubble Ultra Deep Field (HUDF, $\simeq$ 40 kpc $\times$ 80 kpc) and behind the Hubble
Frontier Field galaxy cluster MACSJ0416 ($\simeq 40$kpc), spatially associated with
groups of star-forming galaxies.  VLT/MUSE integral field spectroscopy
reveals a complex structure with a spatially-varying double peaked Ly$\alpha$ emission. 
Overall, the spectral profiles of the two Ly$\alpha$ nebulae are remarkably similar, both
showing a prominent blue emission, more intense and slightly broader than
the red peak.
From the first nebula, located in the HUDF, no X-ray emission has been detected,
disfavoring the possible presence of AGNs.
Spectroscopic redshifts have been derived for 11 galaxies within $2\arcsec$ from the
nebula and spanning the redshift range $1.037<z< 5.97$.
The second nebula, behind MACSJ0416, shows three aligned star-forming galaxies
plausibly associated to the emitting gas. In both systems, the associated galaxies
reveal possible intense rest-frame-optical
nebular emissions lines \oiiidoub\ +H$\beta$ with equivalent widths as high as 
1500\AA~rest-frame and star formation rates ranging from a few to tens of solar
masses per year.
A possible scenario is that of a group of young, star-forming galaxies 
sources of escaping ionising radiation that induce Ly$\alpha$ fluorescence, 
therefore revealing the kinematics of the surrounding gas.
Also Ly$\alpha$ powered by star-formation and/or cooling radiation may resemble the
double peaked spectral properties and the morphology observed here.
If the intense blue emission is associated with inflowing gas, then we may be witnessing
an early phase of galaxy or a proto-cluster (or group) formation.

\end{abstract}


\section{Introduction}

Exchanges of gas between galaxies and the ambient intergalactic
medium play an important role in the formation and evolution of galaxies.
Circumgalactic gas at high redshift ($z>3$) has been observed through
absorption line studies using background sources close to foreground galaxies
\citep[e.g.,][]{lanzetta95, chen01,adelberger03,steidel10,giavalisco11,turner14}. 
The presence of a significant amount of
circumgalactic gas has also been revealed through the detection of
extended Ly$\alpha$ emission at several tens kpc scales around single galaxies 
\citep[$L_{\alpha} \simeq 10^{42} erg s^{-1}$, e.g.,][]{steidel10,caminha16,patricio16,wisotzki16}
and up to hundreds of kpc scale around
QSOs and/or high redshift radio galaxies 
\citep[$L_{\alpha} \simeq 10^{44} erg s^{-1}$, e.g.,][]{borisova16,cantalupo14,swinbank15}.

While the origin of the extended Ly$\alpha$ emission
is still debated, it is clear that the circumgalactic
gas must be at least partly neutral. 
Extended Ly$\alpha$ emission is therefore a viable tool to investigate
the presence, status and dynamics of the surrounding hydrogen gas, from single
galaxies or galaxy groups. Various processes can be investigated, e.g.,
(1) the search for outflowing/inflowing material provide insights about 
feeding mechanism for galaxy formation and regulation on galactic
baryonic/metal budgets and connection with the IGM, and
(2) indirect signature of escaping ionizing radiation that illuminate
inflowing/outflowing neutral hydrogen gas shaping the Ly$\alpha$ emission profile. 
The latter is connected with ionization capabilities of sources on their local environment
\citep{rauch11,rauch16}. 

The extent of Ly$\alpha$ nebula is found to be
strongly related to the luminosity of a central source, with the largest 
nebulae extending to hundreds of kpc around luminous AGN 
\citep[e.g.,][]{borisova16,swinbank15,cantalupo14,hennawi15}.
The shape of these nebulae is often found to be symmetrical or filamentary
around a central source \citep[e.g.,][]{hayes11,wisotzki16,patricio16}, 
where more luminous sources show more circular morphologies. 
The central source is in agreement with
proposed mechanisms responsible for extended emission,
however, there are nebulae where no central source is detected \citep{prescott12}. 

The diverse origin of LABs can also be seen in the dynamics of Ly$\alpha$ nebulae.
Most of the studied LABs have a chaotic distribution of Ly$\alpha$ emission 
\citep[e.g.,][]{christensen04,caminha16,prescott12,patricio16,francis13} 
which is in agreement with Ly$\alpha$ scattering or an ionising central source as the
origin of the extended emission. 
The discovery of rotating LABs \citep[e.g.,][]{prescott15,martin15} 
indicates that also cold accretion flows can be responsible for extended Ly$\alpha$ emission.

In this work we report on two very similar systems at approximately the same redshift ($z=3.3$) 
discovered in two different fields: 
one recently observed with long-list spectroscopy by \citet{rauch11,rauch16} 
in the Hubble Ultra Deep Field (HUDF, hereafter), and a second one discovered as a multiply imaged
system in the Hubble Frontier Fields cluster MACSJ0416.
Both of these systems have been observed with VLT/MUSE integral field spectroscopy, which 
revealed extended Ly$\alpha$ emission 
coincident with a group of star-forming galaxies. 
We present a study of morphology and spectral profile of these systems, possibly tracing
outflowing and inflowing gas.

In the following discussion we assume a flat cosmology with
$\Omega_{m}=0.3$, $\Omega_{\Lambda}=0.7$ and $H_{0}=70 \, {\rm km\, s^{-1}\, Mpc^{-1}}$,
corresponding to 7.6kpc proper for 1\arcsec\ separation at $z=3.3$.

\begin{figure*}
\centering
\includegraphics[width=14cm]{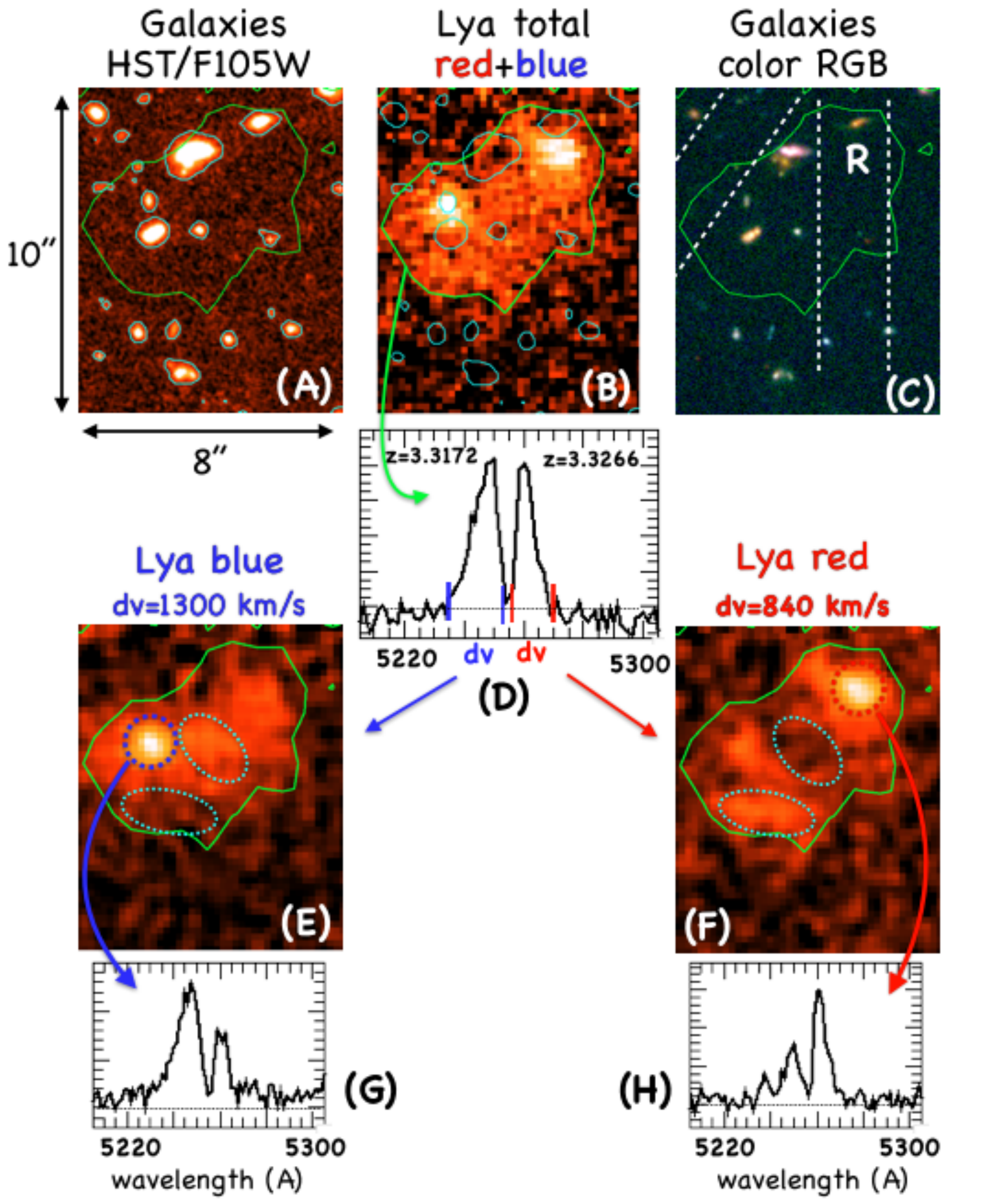}
\caption{Panel {\bf A}: HST F105W-band image of the region covering the Ly$\alpha$ nebula.
Galaxies are highlighted with contours (only for eye-guidance). Panel {\bf B}: Ly$\alpha$ nebula as the sum of the
two Ly$\alpha$ peaks, blue and red. Contours indicate the position of galaxies, the green contour
marks the $Ly\alpha$ emission above 2-sigma from the background. 
Panel {\bf C} shows the color image derived from the HST/ACS F435W, F606W, and z850LP bands.
Panel {\bf D}: summed (red + blue) one-dimensional Ly$\alpha$ spectral profile
integrated within the green contour, where two peaks are evident. 
The Ly$\alpha$ spatial maps of the blue and red components are shown in panels {\bf E}
and {\bf F}), respectively. These are computed by collapsing the signal in the velocity intervals 
dv=1300 and 840 km~s$^{-1}$) (marked with blue and red segments). 
Clearly, the blue and red peaks of the Ly$\alpha$ emission originate from two different, 
spatially separated regions. The Ly$\alpha$ one-dimensional spectra extracted over these two regions
are shown in panels {\bf G} and {\bf H}. These spectra are extracted adopting circular apertures of
$1.2\arcsec$ diameter (blue and red dotted circles). Dotted ellipses mark two regions over which 
each of the peaks of the Ly$\alpha$ emission either dominates or is depressed.}
\label{nebula}
\end{figure*}

\begin{table}
\footnotesize
\caption{List of parameters}
\begin{tabular}{ l r } 
\hline
Ly$\alpha$ nebula - HUDF& {}\\ 
\hline
Right ascension(J2000): & {\bf $03h32m39.0s$}\\
Declination(J2000): & {\bf $-27^{\circ}46^{'}17.0''$}\\
Redshift(blue,red): & 3.3172,3.3266 ($\pm 0.0006$)\\
L(Ly$\alpha$) blue (\ergs): & $(5.5 \pm 0.1)\times10^{42} $\ergs \\
L(Ly$\alpha$) red (\ergs):   &  $(4.0 \pm 0.1) \times10^{42} $\ergs \\
\hline
Possible galaxy counterparts & {}\\ 
\#2-16373(SFR; Mass) & 4\msunyr ; $2.9\times10^{7}$\msun \\ 
\#3-16376(SFR; Mass) & 46\msunyr ; $1.2\times10^{9}$\msun \\ 
\#4-16148(SFR; Mass) & 0.1\msunyr ;$5.5\times10^{8}$\msun \\ 
\#6-16330(SFR; Mass) & 2\msunyr ; $3.2\times10^{7}$\msun \\
\#10-16506(SFR; Mass) & 100\msunyr ; $9.0\times10^{8}$\msun \\ 
\hline \hline
Ly$\alpha$ nebula - MACSJ0416& {}\\ 
\hline
Right ascension(J2000)[A]: & $04h16m10.9s$\\
Declination(J2000)[A]: & $-24^{\circ}04^{'}20.7''$\\
Right ascension(J2000)[B]: & $04h16m09.6s$\\
Declination(J2000)[B]: & $-24^{\circ}03^{'}59.7''$\\
Redshift(blue,red): & 3.2840,3.2928 ($\pm 0.0006$)\\
L(Ly$\alpha$) blue (\ergs): & $(4.4 \pm 0.1)\times10^{42} $\ergs \\
L(Ly$\alpha$) red (\ergs):   &  $(3.7 \pm 0.1) \times10^{42} $\ergs \\
\hline
Galaxy counterparts & {}\\ 
\#1439(SFR; Mass) & 1.5\msunyr ; $4.4\times10^{8}$\msun \\
\#1443(SFR; Mass) & 1.2\msunyr ; $3.0\times10^{9}$\msun \\
\#1485(SFR; Mass) & 3.4\msunyr ; $1.5\times10^{10}$\msun \\
\hline \hline
\end{tabular}
\label{tab:valori}
\end{table}

\section{MUSE integral field spectroscopy}

\subsection{Hubble Ultra Deep Field and Hubble Frontier Fields MACS~J0416}

The MUSE instrument mounted on the VLT \citep{bacon12}.
is an efficient integral filed spectrograph  highly suitable to blindly look for extended line emission. 
Its relatively large field of view (1 arcmin$^{2}$), spectral range (4750-9350\AA), relatively high spatial
(0.2\arcsec) and spectral ($R\sim3000$) resolution,
and stability allowed us to discover extended Ly$\alpha$ emission, down to 
faint flux levels.

MUSE observations (Prog. ID 094.A-0289, P.I. Bacon) of the
Hubble Ultra Deep Field \citep{beckwith06} were obtained between October and December 2014.
Data have been reduced using the standard ESO pipeline version 1.2.1.
The raw calibration and science exposures of each single
night have been processed, and combined into the final data cube
(see Caminha et al. 2016 for a more specific description).
The seeing conditions were good, with an average of $\simeq 0.8$\arcsec, and 81\% of the raw
frames with seeing $<1$\arcsec\ based on the DIMM monitor at Paranal
(the lack of bright stars in the pointing did not allowed us to directly measure
the seeing on MUSE data).
Moreover, a visual inspection of all exposures (50 with DIT=1500s) from the stacked data cubes did
not show evidence of significant variations in observational conditions.
We obtain a reduced datacube centered at  RA=03h32m38s, DEC=-27$^{\circ}$46$'$44$''$
of 19.5 hours total integration time.
We identified a spatially extended Ly$\alpha$ emission $\simeq 8\arcsec \times 4\arcsec$
wide ($\simeq 70\times40$ proper kpc) at $z=3.32$ in the HUDF
centered at coordinates RA=03h32m39.0s, DEC=$-27^{\circ}46'17.0''$. 
The spectral line profile shows two main peaks, named ``red'' and ``blue'',
hereafter, separated by a trough at $z\simeq3.322$.

We also analysed MUSE data obtained on the North-Eastern portion of the Hubble Frontier Fields cluster MACS~J0416 
\citep{lotz14,kokom14}\footnote{http://www.stsci.edu/hst/campaigns/frontier-fields/}.
Data have been taken as part of the GTO program 094.A-0115 (P.I. Richard).

Observations were obtained in November 2014, for a total of 2 hours splitted in 4 exposures. 
Data have been reduced as described above (and we defer the reader to Caminha et al. 2016 for details)
and produced a datacube of 2.0 hours total integration time, centered at RA=04h16m09.95s, DEC=-24$^{\circ}$04$'$01.9$"$
with position angle 45$^{\circ}$. Also in this field we identified a strongly lensed Ly$\alpha$ nebula at $z=3.33$ with 
a size of $\simeq 40$kpc proper with a spectral shape remarkably similar to the one discovered in the HUDF.
Two multiple images of the lensed nebula have been identified at coordinates RA=04h16m10.8s, DEC=-24$^{\circ}$04$'$20.5$"$
and RA=04h16m09.6s, DEC=-24$^{\circ}$04$'$00.0$"$.

In the following we focus on the details of these two systems. The basic properties are reported in Table~\ref{tab:valori}.

\section{Ly$\alpha$ emitting nebula in the Hubble Ultra Deep Field}

\subsection{Previous long-slit spectroscopy}

A portion of the the system studied in this work was observed with extremely deep, 
blind long-slit spectroscopy \citep{rauch11} and, recently, with
additional long slit observations \citep{rauch16}.
Rauch et al. reported a complex structure, i.e., diffuse, fan-like, blue-shifted
Ly$\alpha$ emission and a DLA system in front of galaxy R (see Figure~\ref{nebula}, top-right). 
Various scenarios have been discussed, in particular the Ly$\alpha$ emission 
can be explained if the gas is inflowing along a filament behind the galaxy 'R' 
(see Figure~\ref{nebula}, in which their slit orientations are shown, panel C) and emits
fluorescent Ly$\alpha$ photons induced by the ionizing flux escaping from the
galaxy. 

Long-slit spectroscopy, however,  probe only a small volume and
offer a marginal/limited view of the whole system.  Our study benefit from
a number of other key improvements over previous works: (i) the large field
of view and spectral resolution of the MUSE instrument is essential
for capturing many galaxies simultaneously in a single ultradeep
pointing; (ii) deep HST imaging observations cover the full
MUSE field of view; (iii) homogeneous sub-arcsec spatial resolution
for all objects is available.
In the following we describe the system as observed with MUSE.

\subsection{Ly$\alpha$ nebula properties}

Using the MUSE datacube we constructed ``pseudo-narrowband''
(NB) images of the extended emission, centered on the position and
wavelength of the corresponding Ly$\alpha$ line.

Before extracting the NB images we performed an additional preprocessing
as described in the following. At the low flux levels of interest in this study,
source crowding becomes a serious issue for many objects.
Several close neighbours within a few arcsec in projection are visible in the HST data. 
Since these neighbours are typically foreground sources, 
they contaminate the NB signal only with their continuum emission.
A conventional method to remove the contaminating continuum is to subtract
a suitably scaled off-band image. We adopted an similar method that fully
exploits the information contained in the MUSE datacube: we
first median-filtered the datacube in the spectral direction with
a very wide filter window of $\pm 100$ spectral pixels; this produced
a continuum-only cube with all line emission removed and with
the continuum spectra of real objects being heavily smoothed.
The Ly$\alpha$ image has been computed by 
choosing the band limits such that about
95\% of the total line fluxes were included. Being the Ly$\alpha$ nebula double peaked,
the bandwidths resulted to be
14 and 9 spectral pixels (1pixel = 1.25\AA) for the blue and red Ly$\alpha$ peaks, respectively,
corresponding to velocity width of 1300\kms\ and 840\kms.
We then subtracted this filtered cube from the original data Ly$\alpha$ image and
thus obtained an essentially pure emission line cube which was
(to first order) free from any continuum signal. 
As an example the $z=1.037$ foreground galaxy has been optimally
removed (see below).
The spatial maps of the red, blue and ``red+blue'' Ly$\alpha$ nebulae are
shown in Figure~\ref{nebula}. 
The integrated Ly$\alpha$ flux computed within a polygonal aperture defined 
following the green contour of Figure~\ref{nebula} is $(6.1 \pm 0.1) \times 10^{-17}$\ergscm\ and
$(4.4 \pm 0.1) \times 10^{-17}$\ergscm, corresponding to $5.5 \times 10^{42}$\ergs\ and 
 $4.0 \times 10^{42}$\ergs\, for the blue and red components, respectively.
 
\section{Results}

\subsection{Ly$\alpha$ emitting nebula in the UDF}

\begin{figure*}
\centering
\includegraphics[width=14cm]{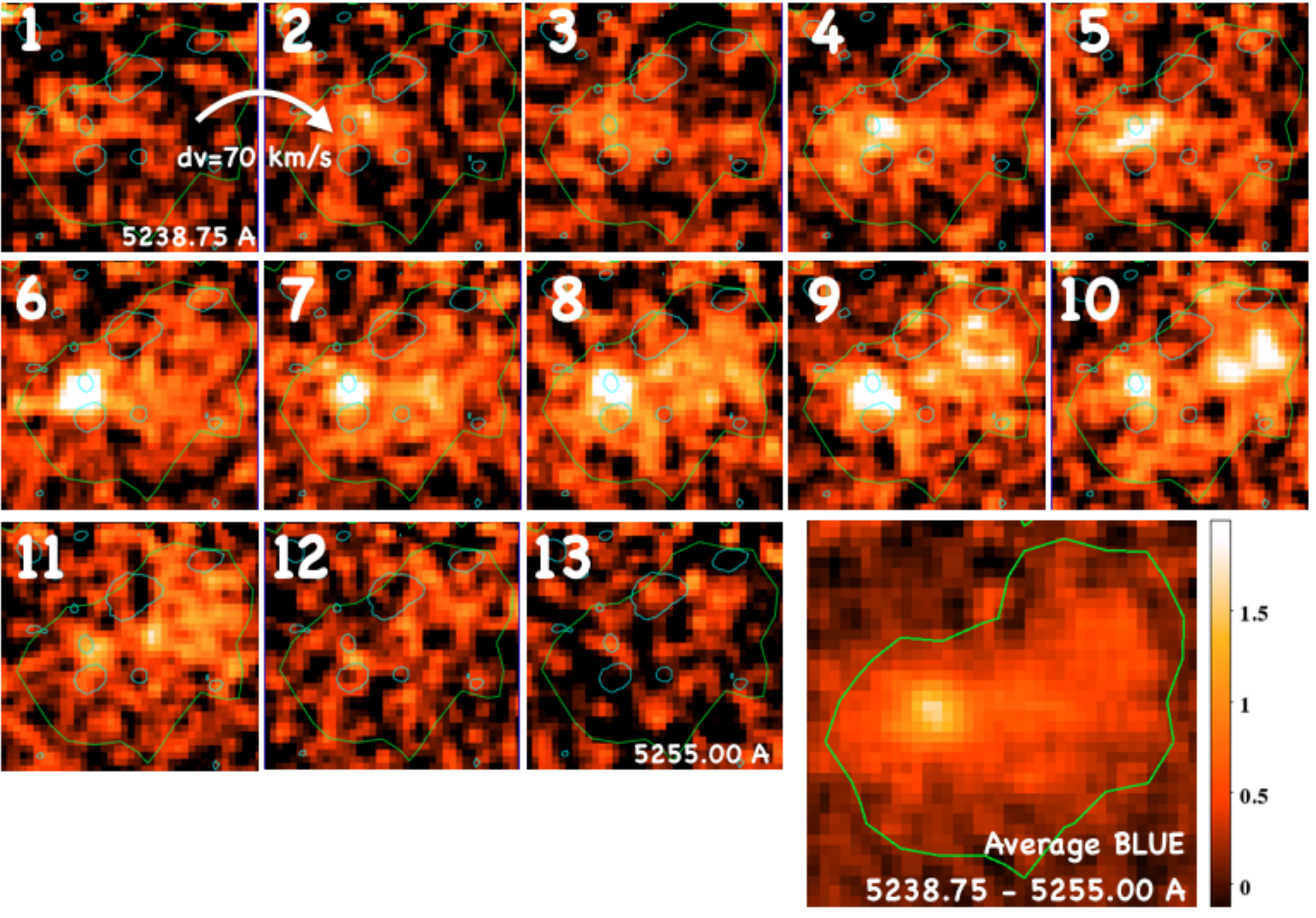} 
\caption{Unpacked images of the blue Ly$\alpha$ component in 13 slices of 1.25\AA~each.
The images are sorted by progressively increasing wavelength from top-left to bottom-right
spanning the range 5238.75\AA\ -- 5255.00\AA. 
Cyan contours highlight the position of galaxies as shown in 
Figure~\ref{nebula}. A dominant emission
is present on the left (East) with an arising plume toward the right-side (West) as wavelength increases. 
In the bottom-right panel the average of the all slices is shown. The green line indicates the Ly$\alpha$ 
nebula contour from the ``red+blue'' map. The color coded bar indicate units of $10^{-20} erg^{-1} s^{-1} cm^{-2} pix^{-2}$,
with 1pix=0.2\arcsec.}
\label{lyablue}
\end{figure*} 

When examining the Ly$\alpha$ emission of this nebula both spatially and spectrally, we identified two 
main characteristics:

(1) a double-peaked profile in the spectral domain, with blue and red components peaked
at redshift 3.3172 and 3.3266 (and $dv=650$\kms), each one showing a
blue/red tail. The bluer peak is more intense and broader than the red one
(FWHM $\sim 520\pm50$\kms\ and $280\pm50$\kms, respectively).

(2) a different overall spatial distribution of the
blue and red emission. From the corresponding spatial maps 
it is evident  that the bulk of the emission of the two peaks are
segregated (marked with blue/red
dotted circles of $1.2\arcsec$ diameter in Figure~\ref{nebula}). Again, the spectral
shape of these two selected regions follows the profiles described above
(the extracted one-dimensional spectra from the circular apertures 
are shown in the same figure),
in which the strongest blue emission is also broader than the red one.

\begin{figure*}
\centering
\includegraphics[width=14cm]{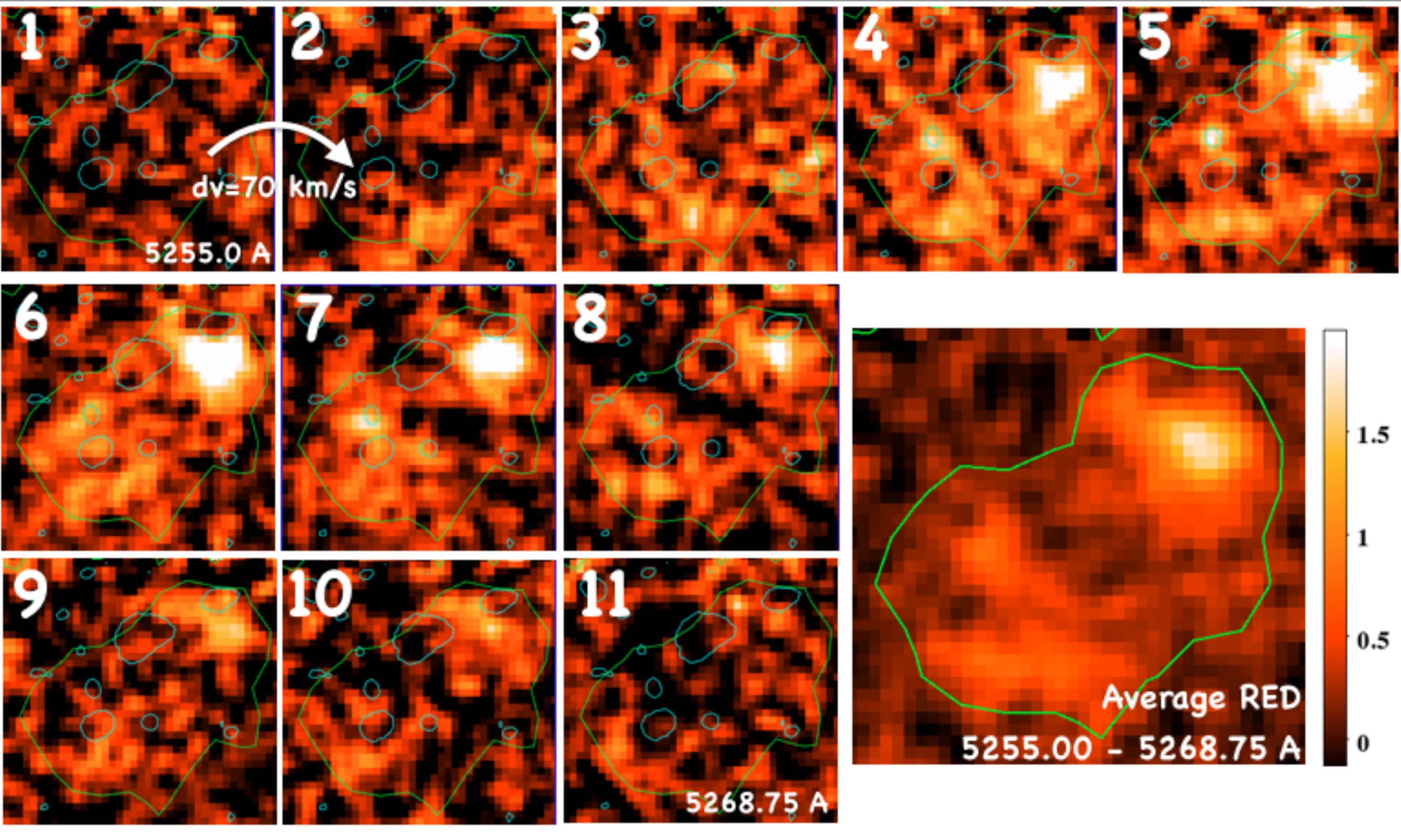} 
\caption{Unpacked images of the red Ly$\alpha$ component in 11 slices of 1.25\AA~each. 
The images are sorted by progressively increasing wavelength from top-left to bottom-right
spanning the range 5255.00\AA\ -- 5268.75\AA. 
Cyan contours highlight the position of galaxies as shown in 
Figure~\ref{nebula}.  A dominant emission
is present on the right (West) with structure extending toward the left side (East). In the bottom-right
panel the average of the all the slices is shown. The green line indicates the Ly$\alpha$ nebula contour from
the "red+blue" map. The color coded bar indicate units of $10^{-20} erg^{-1} s^{-1} cm^{-2} pix^{-2}$,
with 1pix=0.2\arcsec.}
\label{lyared}
\end{figure*}

The ratio of the two peaks depends on the aperture adopted and the 
position on the nebula. For example, we identified two regions  
that show a deficit either of blue or red Ly$\alpha$ emission. These are marked with dotted ellipses in 
Figure~\ref{nebula} (panel E and F).
Further details of the spatial behavior of the red and blue emissions along their spectral
direction is shown in Figures~\ref{lyablue} and ~\ref{lyared}, in which each slice
with step of 1.25\AA~is shown ($dv=70$\kms).
In particular the blue emission has been expanded in 13 slices from 5238.75\AA~to 5253.75\AA\,
and appears to grow first from the East side (where it is dominant, e.g., slice 4) and subsequently
moves toward the West side following a non-uniform pattern, but maintaining a blue
plume over the entire region (slices 10-11). 
The blue emission covers also regions where the red one is present.
The average of all the slices within dv$\simeq1300$\kms\ is shown in Figure~\ref{lyablue}.

The red emission has also been expanded in 11 slices from 5255.0\AA~to 5267.5\AA\ 
and is mainly located where its main peak has been identified (as discussed above),
however in few slices it shows structures that extend toward the regions where the blue
emission is dominant. The average of all the slices within dv$\simeq840$\kms\ is shown in Figure~\ref{lyared}.

\begin{figure*}
\centering
\includegraphics[width=14cm]{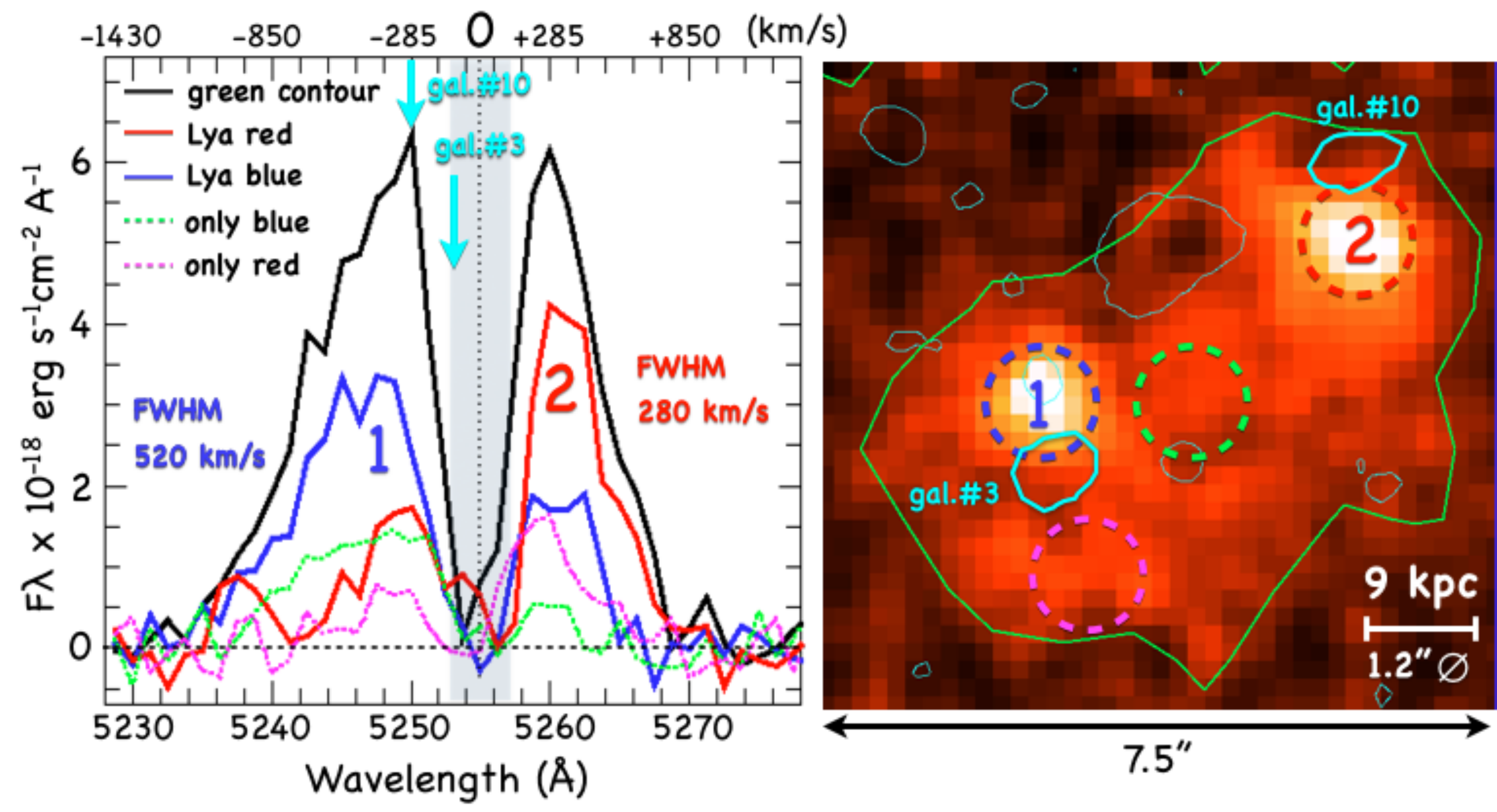} 
\caption{{\bf Right Panel:} the position of four circular apertures ($1.2\arcsec$ diameter) are shown
on the image of the Ly$\alpha$ nebula (the image shown here is "red+blue").
Green and cyan lines mark the contour of the nebula and
galaxies (as described in Figure~\ref{nebula}). The two apertures centered on the dominant emissions (1 and 2) are
shown in blue and red on the left panel. The galaxy LBG\#3 is marked with thick cyan contour and its velocity
position relative to the nebula is shown in the left panel. {\bf Left Panel:} The one-dimensional Ly$\alpha$ profiles
extracted from the apertures indicated on the right panel. In particular the total profile (extracted from the
green contour) is shown in black, and the spectra arising from the apertures 1 and 2 on top of the main emissions
are shown in blue and red, respectively (the FWHM are also indicated).
Dotted magenta and green lines mark the Ly$\alpha$ profiles in regions where the blue and red emission
are deficient, respectively.
In general, the double peaked Ly$\alpha$ emission persist over the nebula with the position of the trough
consistent with the marked gray region, $\simeq 230$\kms\ wide. }
\label{lyaprofile}
\end{figure*}

We map the Ly$\alpha$ emission spatially by extracting spectra from apertures covering
different regions of the nebula. In particular the one-dimensional profiles are shown in
Figure~\ref{lyaprofile}, where
the two main emissions are shown with blue and red colors. Again, this clearly shows a 
broader blue component and the different intensities among the two red and blue peaks. 
It is worth noting that the nebula shows a double peaked Ly$\alpha$ emission from both positions,
1 and 2, marking the blue and red dominant emissions (see Figure~\ref{lyaprofile}).
The wavelength location of the Ly$\alpha$ trough is not changing significantly, 
remaining within d$\lambda \simeq 4$\AA\ (dv$\simeq 230$\kms) from the central position (5255\AA, $z=3.3226$),
derived by integrating over the entire nebula (green contour in Figure~\ref{lyaprofile}).

\subsubsection{The broad-band counterparts}

\begin{figure*}
\centering
\includegraphics[width=16cm]{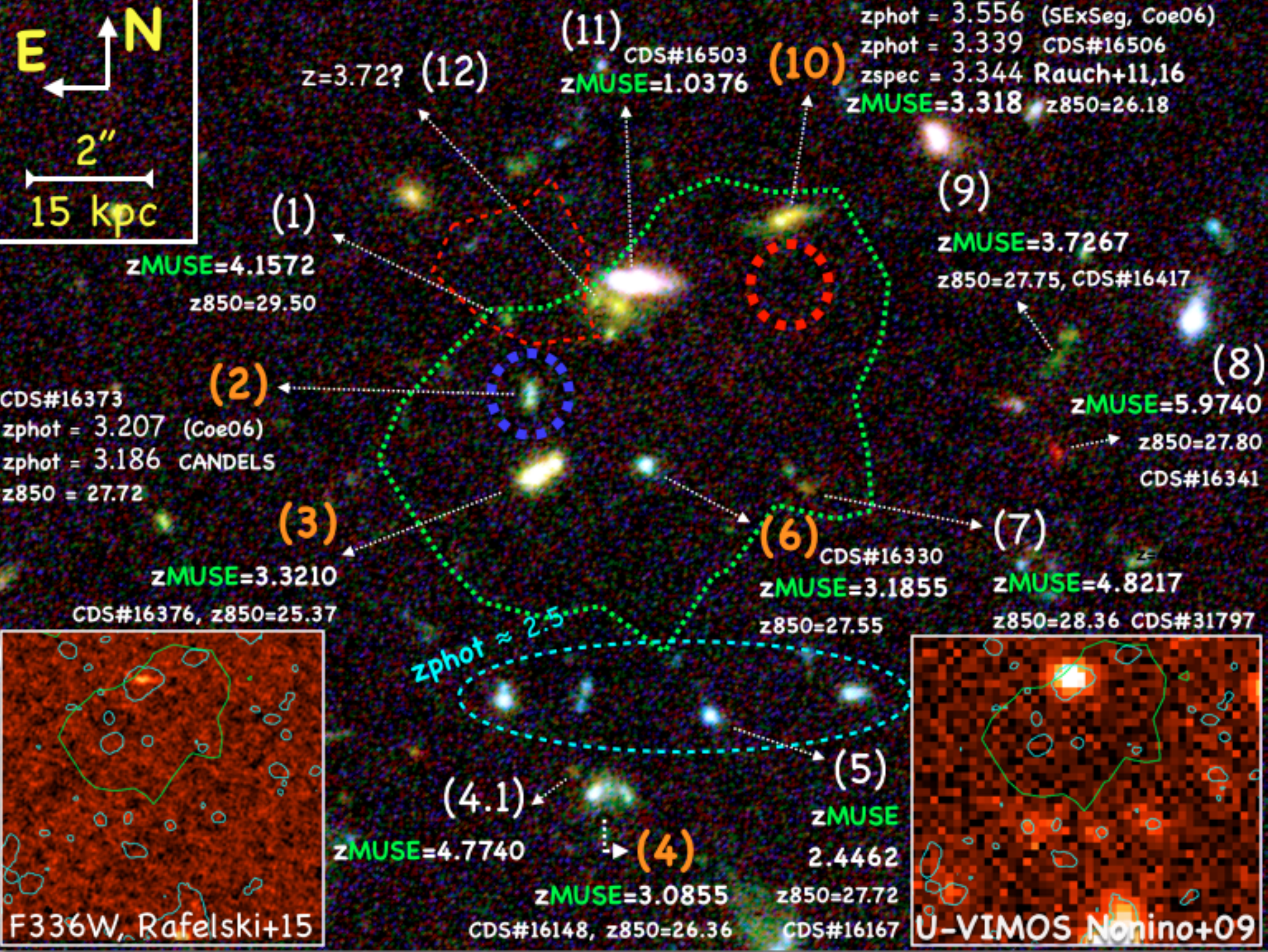} 
\caption{Galaxies surrounding the $Ly\alpha$ nebula (dotted green contour) are shown superimposed to the HST color image
centered on the nebula (F435W, F606W and i775W, Beckwith et al. 2006). 
Dashed red and blue circles mark the positions of the dominant Ly$\alpha$ red and blue emissions, respectively
(see also Figure~\ref{nebula}, panel B).
Redshifts extracted from MUSE
are highlighted in green and $z_{850}$ magnitudes derived from \citet{coe06}. The CANDELS identifier
is also reported (CDS\#) for relevant sources (in particular the IDs highlighted in orange are those possibly
associated to the system, see text).
The photometric redshifts are reported, if no reliable features were found in the MUSE datacube
(\citealt{coe06} and CANDELS).
In the bottom-left/right
the same region of the sky is shown in the deep F336W/U-VIMOS bands. This highlights the drop of the majority of the
sources considered in this study, therefore supporting their high redshift nature, $z\gtrsim3$. Galaxies close to the nebula
are marked in orange, in particular galaxy ID=3 lies in the Ly$\alpha$ trough (see text).
The thin-dotted red contour marks a possible Ly$\alpha$ halo at $z=3.723$, shown in the bottom left panel of
 Figure~\ref{musespectra}.}
\label{broadband}
\end{figure*}

\begin{figure*}
\centering
\includegraphics[width=16cm]{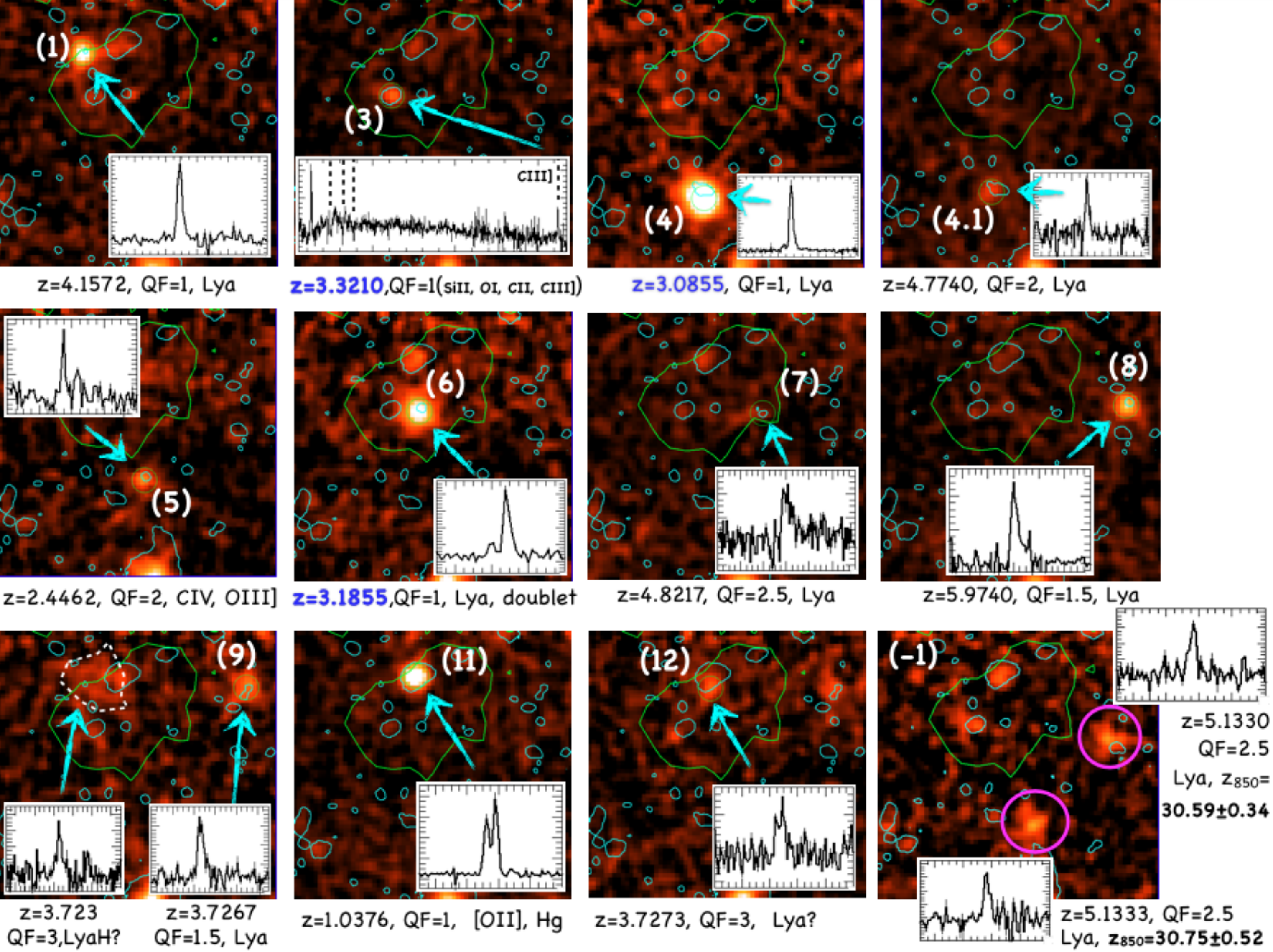} 
\caption{Snapshots of the MUSE spectra ($d\lambda=1.25$\AA) of the identified lines.
The IDs correspond to the galaxies marked in Figure~\ref{broadband} following from top-left to
bottom-right the counterclockwise order. Redshifts are
reported in the bottom of each image, with the quality (QF, 1=secure, 2=plausible, 3=tentative) and the main spectral
feature identified. In the bottom-right, two emission lines not associate to any CANDELS source are reported.
They are consistent with the position of two sources identified by \citet{coe06}, with magnitudes $z_{850} \gtrsim 30.5$.}
\label{musespectra}
\end{figure*}

We discussed above about the spatial distribution and spectral profile of the emitting Ly$\alpha$
nebula. Here we discuss the identification of possible broad-band counterparts.
MUSE integral field spectroscopy allows to extract a $\sim 20$hr spectrum
of each of the galaxies detected in the HUDF down to the limiting depth. We were able to measure redshifts
for 11 galaxies in the region surrounding the nebula, two of them - marked as \#7 and \#12 - 
with only tentative redshift estimates (see Figure~\ref{broadband}). The MUSE spectra 
are shown in Figure~\ref{musespectra}, where the wavelength slices corresponding
to the peak of the most prominent emission line are shown.
The redshifts of these galaxies are in the range $1<z< 6$.
No X-ray emission has been detected from the 7Ms Chandra data, neither from the nebula nor
from any of the galaxies close by, and no evident high ionization emission lines 
(e.g., \civ, \heii) have been identified.

Several other galaxies show colors and photometric redshifts
(from \citet{coe06} and CANDELS \footnote{https://rainbowx.fis.ucm.es/Rainbow\_navigator\_public/})
consistent with the redshift of the nebula and are reported in Figure~\ref{nebula}.  
Deep VLT/VIMOS U-band \citep{nonino09} and HST/F336W \citep{rafelski15} images
are shown in Figure~\ref{broadband}.
The dropout in these bands of several galaxies further support their high redshift nature.
We confirm the presence of a foreground galaxy at $z=1.037$ and find five additional sources at $z>3.7$,
which are certainly not associated with the nebula.
The position of a group of galaxies at zphot$\simeq 2.5$
is also marked in the figure. One of them is spectroscopically confirmed at $z=2.4462$.
As expected, they are also detected in the VIMOS U-band.
Three galaxies (namely,  \#3, \#4 and \#6) have spectroscopic redshifts within dz=0.23 
(corresponding to dv$<$16000\kms) from 
the red or blue Ly$\alpha$ emissions (all redshifts are reported in Figure~\ref{broadband}
and ~\ref{musespectra}).

\begin{figure*}
\centering
\includegraphics[width=14cm]{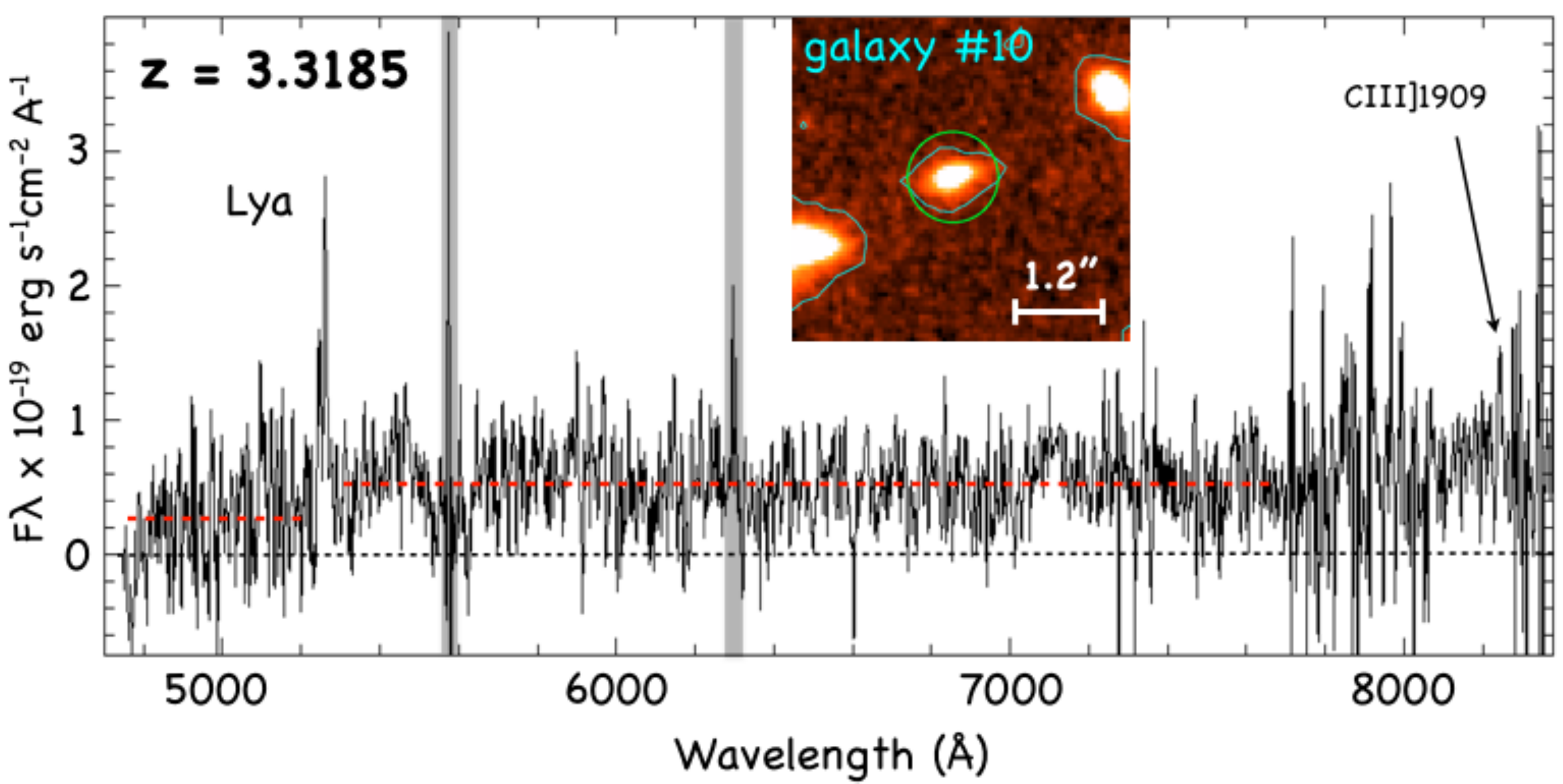} 
\caption{The MUSE spectrum of galaxy \#10 calculated within a 1.2\arcsec diameter aperture (green
in the inset)  is shown. While no evident absorption lines have been detected, the
possible \ciiidoub\ in emission and the clear continuum break (marked with dashed red line) are fully compatible
with the Ly$\alpha$ emission of the nebula (indicated as Ly$\alpha$ in the figure. The resulting cross correlation
produces a redshift z=3.3185.}
\label{gal10}
\end{figure*}

We were not able to confirm the redshift reported by \citep{rauch11,rauch16} for the galaxy \#10, 
at $z=3.344$, in particular we do not detect the \heii\ emission line they reported.
The MUSE spectrum in the expected position of \heii\ is free from sky lines
(see Figure~\ref{gal10}), and does not show evident line emission down to $3\times10^{-18}$\ergscm\ 
at 3-sigma level.
While the galaxy is a clear U-band dropout with photometric redshifts 3.556 \citep{coe06}. 
and 3.339 (CANDELS), and consistent with the other galaxies and nebula discussed here, 
we consider the redshift still uncertain. Even adopting the redshift
reported by \citep{rauch16}, the galaxy would be at 1200\kms\, therefore a possible Ly$\alpha$ 
emission line would appear redward of the reddest peak of the nebula.

We noted another galaxy presenting a multi-blob morphology, close ($<0.3\arcsec$) to the $z=1.037$
foreground galaxy and toward the south-east. This object is not recovered by any public photometric catalog, 
neither CANDELS \citep{guo13} nor GOODS-ACS \citep{giava04}.
Additionally, this ``multi-blob'' object is a U-band dropout, as can be inferred from the high spatial resolution 
deep HST/F336W-band imaging, therefore it is plausibly at $z\gtrsim3$. 
We tentatively identified an emission line at 5747\AA\ that, if interpreted as Ly$\alpha$, would place the source at $z=3.727$.
Interestingly, the emission appears spatially resolved suggesting it could be another Ly$\alpha$ blob behind the z=3.2 nebula
discussed here (this putative blob is marked with red dotted line in Figures~\ref{broadband} and ~\ref{musespectra}). Another galaxy
(\#8) has been confirmed at the same redshift (z=3.7267) of this second blob. 
When focusing on this z=3.7 emission superimposed to the HUDF nebula, another Ly$\alpha$ nebula emerges
at 11\arcsec\ from it, in the south-east at $z=3.7123$ with extension $30\times20$kpc physical.
We do not discuss further this projected higher redshift Ly$\alpha$ halos, neither the second system at 11\arcsec,
and focus on the $z=3.2$ nebula. 

The two-dimensional MUSE snapshots (slices of 1.25\AA\ bin) at the wavelength of the main peak of the identified lines are shown in Figure~\ref{musespectra},
together with their one-dimensional zoomed spectra. The reported IDs are the same of Figure~\ref{broadband}. 
In general, line emissions are evident and the spectral resolution ($R \simeq 1800$) is sufficient to detect the low
redshift \oiidoub\ doublet (e.g., galaxy \#11, other than $H\gamma$) and the asymmetric profile typical of the high-z Ly$\alpha$ lines.
Only for one galaxy (\#3, the brightest with $z_{850}=25.37$) the continuum and several absorption lines have been identified together with
the \ciiidoub\ emission (this emission is shown in Figure~\ref{musespectra}). 
Interestingly, two more line emitters have been identified at the same wavelength and associated to two
faint galaxies identified by \citet{coe06} with magnitude $z_{850}=30.59 \pm 0.34$ and 
$30.75\pm0.52$ (detected at 2 and 3 sigma, respectively; they are shown in the bottom-right panel
of  Figure~\ref{musespectra}).
The lines are reasonably well detected (S/N$\sim5-7$). One of the two is possibly asymmetric resembling
the typical Ly$\alpha$ shape. If the lines are Ly$\alpha$ at $z=5.133$, they are obviously not associated with the
system studied in this work. However, the detection of such a faint spectral feature from galaxies that are so faint 
to be barely detected even at the HUDF depth, further supports the capability of the MUSE instrument of 
detecting possible faint counterparts also at the redshift of the nebula. 

\begin{figure}
\centering
\includegraphics[width=8.5cm]{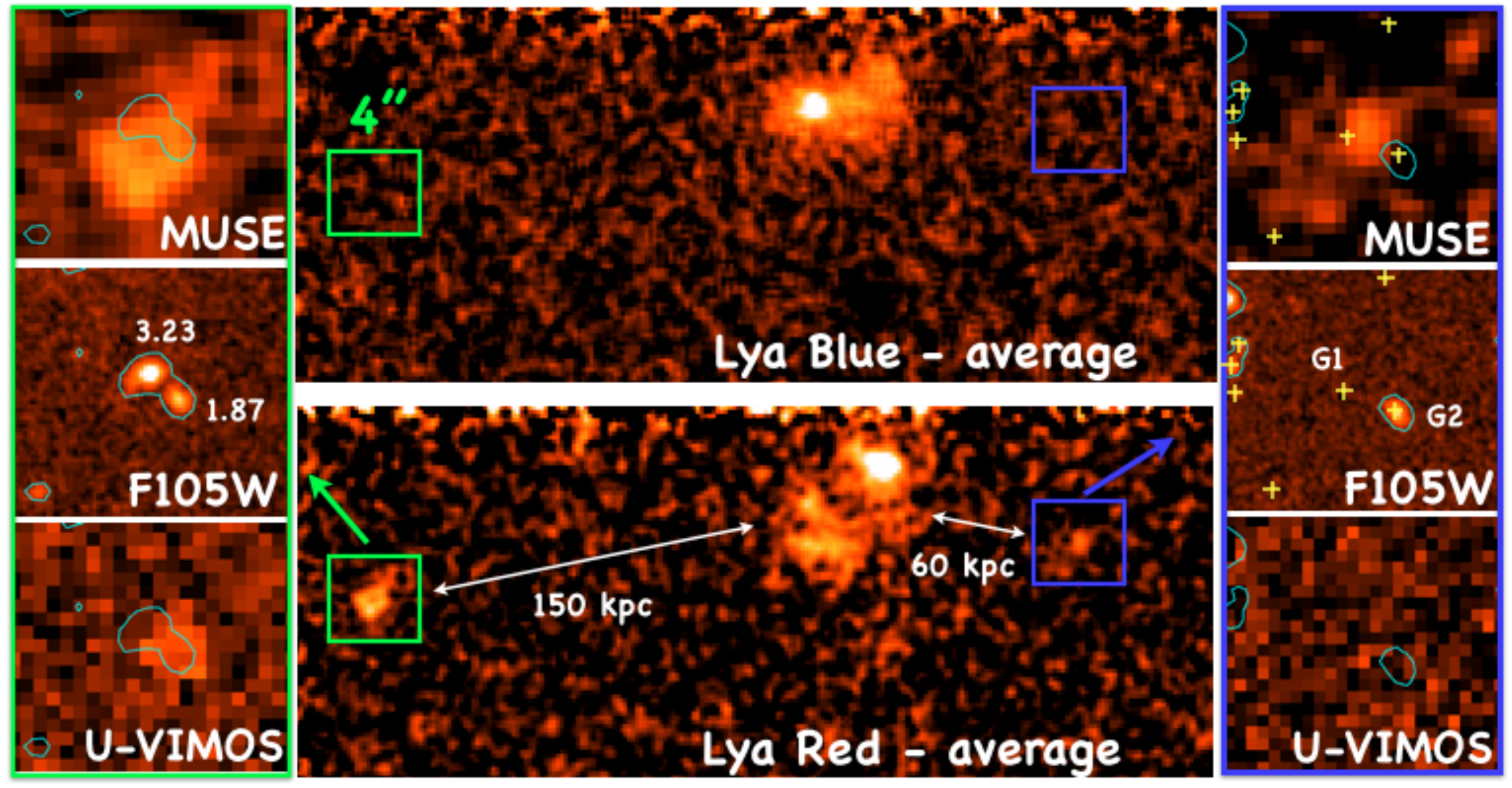} 
\caption{The position of additional Ly$\alpha$ emissions placed at a transverse distance of 150 kpc and 60 kpc proper from the 
nebula are shown. In the middle top/bottom panels the blu/red emission of the nebula averaged within the velocity
slices reported in Figure~\ref{nebula}  are shown with marked the additional Ly$\alpha$ emissions. 
The emissions marked with a green and blue squares at the 
same wavelength of the red Ly$\alpha$ nebula are evident. In the left and right panels a zoom of these emissions are
shown: from top to bottom, the zoomed Ly$\alpha$
emission, the F150W image (with indicated the photometric redshifts from \citealt{coe06}) 
and the deep U-VIMOS image are shown. Cyan contours mark the position of the galaxies as indicated in
 Figure~\ref{nebula}. In the left panels a galaxy with photometric redshift consistent with the Ly$\alpha$ emission
 has been identified (zphot=3.23), while the second lower-z object (zphot=1.87) is also detected in the VIMOS U-band (bottom-left image).
 In the right panels the positions of sources detected in the \citet{coe06} catalog 
 (yellow crosses) are also shown. In particular galaxy G1 and G2 have photometric redshifts $z\simeq 3$ and
 magnitude in the $z_{850}$-band of 30.80 (S/N=2.8) and 28.81 (S/N=14), respectively.}
\label{isolated}
\end{figure}

It is worth noting other two Ly$\alpha$ emissions identified at $\simeq 150$ kpc and 60 kpc proper from the nebula and at the
same redshift of the red Ly$\alpha$ component. Figure~\ref{isolated} shows the positions of these emissions compared with
the position of the underlying galaxies, whose photometric redshift is also reported. The se additional emission lines are
consistent with the photometric redshifts of the closest galaxies, whose magnitudes ranges between 27 up to 30.8 in the
$z_{850}$ band. The Ly$\alpha$ emission appears slightly spatially offset
from the reported galaxies, possibly due to radiative transfer effects (see Figure~\ref{isolated}). 

In the following we focus on the closest galaxies possibly related to the Ly$\alpha$ system.

\subsubsection{The dominant blue and red Ly$\alpha$ emissions and galaxy counterparts}

Three galaxies have been spectroscopically confirmed at redshift
3.3210 (\#3), 3.1855 (\#6) and 3.0855 (\#4) at $<4\arcsec$ from the main blue emission of the nebula,
corresponding to few proper kpc transverse (1\arcsec\ = 7.6kpc proper). 
Another galaxy, \#10, has been plausibly confirmed at redshift 3.3185 close to the red emission.
We describe them more in detail in the following.

\begin{figure*}
\centering
\includegraphics[width=16cm]{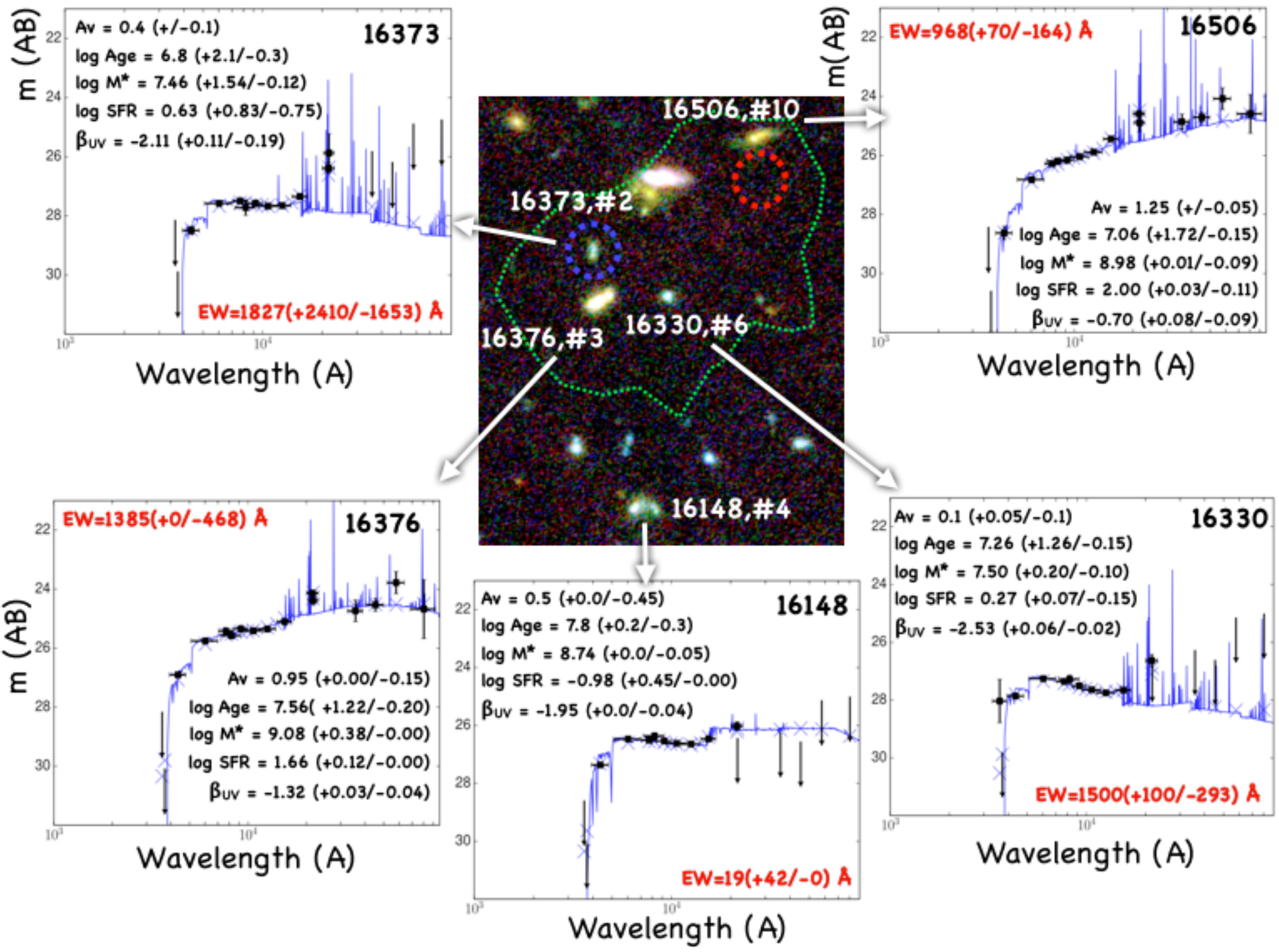} 
\caption{The SED-fitting of the galaxies described in the text and possibly associated to the nebula
are shown. The IDs are same used in the previous figures and Table~\ref{tab:valori}
Basic physical quantities are reported in black, while the estimated equivalent width of the
group \oiiidoub\ and H$\beta$ is reported in red. Apart from galaxy \#4, all the other galaxies show very high
equivalent widths. In particular, galaxy \#4 is the bluest one of the sample, consistent with a negligible
dust attenuation}.
\label{SED}
\end{figure*}

\begin{enumerate}
\item{{\bf Galaxy \#3:} Given its redshift ($z=3.3210$), this is the galaxy that is closest to 
the nebula in the velocity space with a dz=0.0016 from the estimated Ly$\alpha$ trough, 
corresponding to $dv \simeq 110$\kms in the rest-frame of the nebula.
This galaxy is spatially resolved and shows a well detected continuum with
several ultraviolet low-ionization absorption lines in the MUSE spectrum, 
e.g., \siii, \oi\ and \cii\ (see Figures~\ref{broadband} and ~\ref{musespectra}). The estimated stellar mass is
$1.2\times10^{9}$\msun\ with  a star formation rate of 46\msunyr\ (the resulting physical parameters are 
reported in Figure~\ref{SED}).
The \cii\ emission has also been detected and provides an estimate of the systemic
redshift that is fully compatible with the trough of the Ly$\alpha$ nebula (see Figure~\ref{lyaprofile}).
Given the redshift and the angular separation from the main blue Ly$\alpha$ peak (1\arcsec), 
this galaxy is the closest spectroscopically-confirmed source located at a few
tens of physical kpc from the Ly$\alpha$ system.
Interestingly, this galaxy is well detected in the HAWKI-Ks band \citep{fontana14} with 
a magnitude 24.13 at S/N$\simeq 40$, about 1.0 and 0.6 magnitudes brighter than the
adjacent HST/F160W and Sptitzer/IRAC 3.6$\mu m$ bands with magnitudes 25.10 and 24.73
detected at $S/N\gtrsim 40$ and $S/N\simeq 3$, respectively. This discontinuity in the Ks-band
strongly suggests the presence of intense \oiiidoub\ and $H\beta$ nebular emission
lines (see the best SED fitting in Figure~\ref{SED}). Indeed, the result from the SED fitting including 
nebular prescription produces the best solution with a total equivalent width of
EW(\oiiidoub\ + H$\beta$)=1000\AA. Interestingly, the relatively weak 
interstellar absorption lines (especially \cii) the \ciii\ emission, and the strong 
optical rest-frame Oxygen emissions traced by the broad-band Ks magnitude, 
resemble the properties recently identified by \citet{vanzella16} in a Lyman continuum
emitter at similar redshift $z=3.2$ (see also \citealt{debarros16}). We discuss below 
the possible link between escaping ionizing radiation and the surrounding gas.}

\item{{\bf Galaxy \#6:} this is a rather faint ($z_{850}$=27.55), compact, and low-mass
galaxy with stellar mass $3.2\times10^{7}$\msun\ 
showing an extremely blue ultraviolet slope ($\beta=-2.55$).
The SED fitting suggest a low dust extinction, $A_{V}=0.1$,
a SFR=0.3\msunyr. The prominent Ly$\alpha$ emission 
is consistent with the young ages inferred for the burst ($10^{7}$yr). The Ly$\alpha$ line also shows 
a relatively narrow double peaked separation ($dv \simeq 400$\kms) if compared to the
brighter $L^{\star}$ galaxy counterparts of Ly$\alpha$ nebulae \citep[e.g.,][]{kulas12}.
The small peak separation observed in this source is more in line with recent findings at fainter 
luminosities, where optically-thin systems have been identified \citep{vanzella16,karman16}. 
Given its measured redshift ($z=3.1855$), this galaxy might be located in front of the Ly$\alpha$ system.
Interestingly, galaxy \#6 is positioned over the region where the red emission is
minimal, while the blue still survives (see dotted ellipse in Figure~\ref{nebula}).}

\item{{\bf Galaxy \#4:}  this is another galaxy with prominent Ly$\alpha$ emission at 
$z=3.0855$ (see Figure~\ref{musespectra}), identified toward the south, at $<4\arcsec$ from the nebula. 
The inferred stellar mass is $5.5\times10^{8}$\msun\ with a SFR$\simeq 0.1$\msunyr\ (see Figure~\ref{SED}).

The redshift difference between the galaxy and the Ly$\alpha$ trough is dz = -0.237 and corresponds to
$-16400$\kms. Therefore this galaxy appears to be located in front of the system.}

\item{{\bf Galaxy \#2:} 
It is worth noting that
the blue Ly$\alpha$ peak is exactly aligned with another faint galaxy with magnitude 
$z_{850}=27.72$ (galaxy \#2, marked in Figure~\ref{broadband}), for which no spectroscopic redshift
has been measured from MUSE, unless the strong blue Ly$\alpha$ emission of the nebula contains
also the Ly$\alpha$ of the galaxy. Unfortunately it is not possible to clarify this issue 
with the current data. The photometric redshift is, however, very close to the
redshift of the nebula and the estimated stellar mass is  $3\times10^{7}$\msun\ and young age
$\simeq 10^{7-8}$ years, with a relatively blue ultraviolet slope (see Figure~\ref{SED}).
As in the case of galaxy \#3, discussed above, this galaxy has also been detected in the Ks-band (with S/N=6) and
it appears one magnitude fainter in the F160W band, while only upper limits are available in the
IRAC/3.6$\mu m$ and 4.5$\mu m$ channels $m>26$. The presence of intense nebular emission lines 
(\oiiidoub\ and H$\beta$) is, therefore, expected also in this galaxy. When quantified
through the SED fitting, the nebular emission lines should have EW=1800\AA\ (see Figure~\ref{SED}). 
While a confirmation 
would need dedicated near infrared spectroscopy, such strong nebular emission lines are suggestive of high 
ionization parameters and possible Lyman continuum leakage \citep[e.g.,][]{debarros16}.}

\item{{\bf Galaxy \#10:} 
The main red peak of the nebula has no obvious aligned counterparts.
As already discussed above, the closest galaxy is that reported by \citep{rauch11,rauch16}, here
identified as galaxy \#10, for which we were not able to confirm Rauch et al. redshift,
based on \heii\ detection at $z=3.344$ ($\lambda \simeq 7124$\AA). Figure~\ref{gal10} shown the
MUSE spectrum where no clear emission lines have been detected at $\lambda \simeq 7124\pm30$\AA\
down to $\simeq 1\times 10^{-18}$\ergscm\  at 3-sigma level. The position of the \heii\ line is marked with a red
arrow in Figure~\ref{gal10} and is free from sky emission lines. Rather, looking more carefully at he MUSE spectrum a continuum break is 
detected across the Ly$\alpha$ emission. It is not clear if the Ly$\alpha$ (double peaked) emission is 
coming from the galaxy or is the effect of the seeing that spread the signal into the adopted circular aperture
when extracting the MUSE spectrum (marked in green in Figure~\ref{gal10}). 
The cross correlation with typical LBG templates \citep[e.g.,][]{vanzella09} including/excluding the Ly$\alpha$ line
produces the following solutions, z=3.3185/3.3127, respectively (the z=3.3185 would also be compatible with
possible \ciii\ line emission). In both cases the redshift is  slightly lower than
what reported by Rauch et al. and place the galaxy at -690/-285\kms\ from the Ly$\alpha$ trough of the nebula, while the
previous redshift, z=3.344, would put it at +1480\kms, even redder than the red peak of the nebula. 
As discussed by \citet{rauch11,rauch16} this galaxy together with the other counterparts is probably
playing a role in shaping the profile and morphology of the Ly$\alpha$ nebula.
We performed SED fitting also for this galaxy, obtaining a remarkable star formation 
activity ($\sim 100$\msunyr)  with a stellar mass of $\simeq 10^{9}$\msun\ (the more massive
among the counterparts) and relatively red ultraviolet slope (Figure~\ref{SED}).}

\end{enumerate}

\begin{figure*}
\centering
\includegraphics[width=16cm]{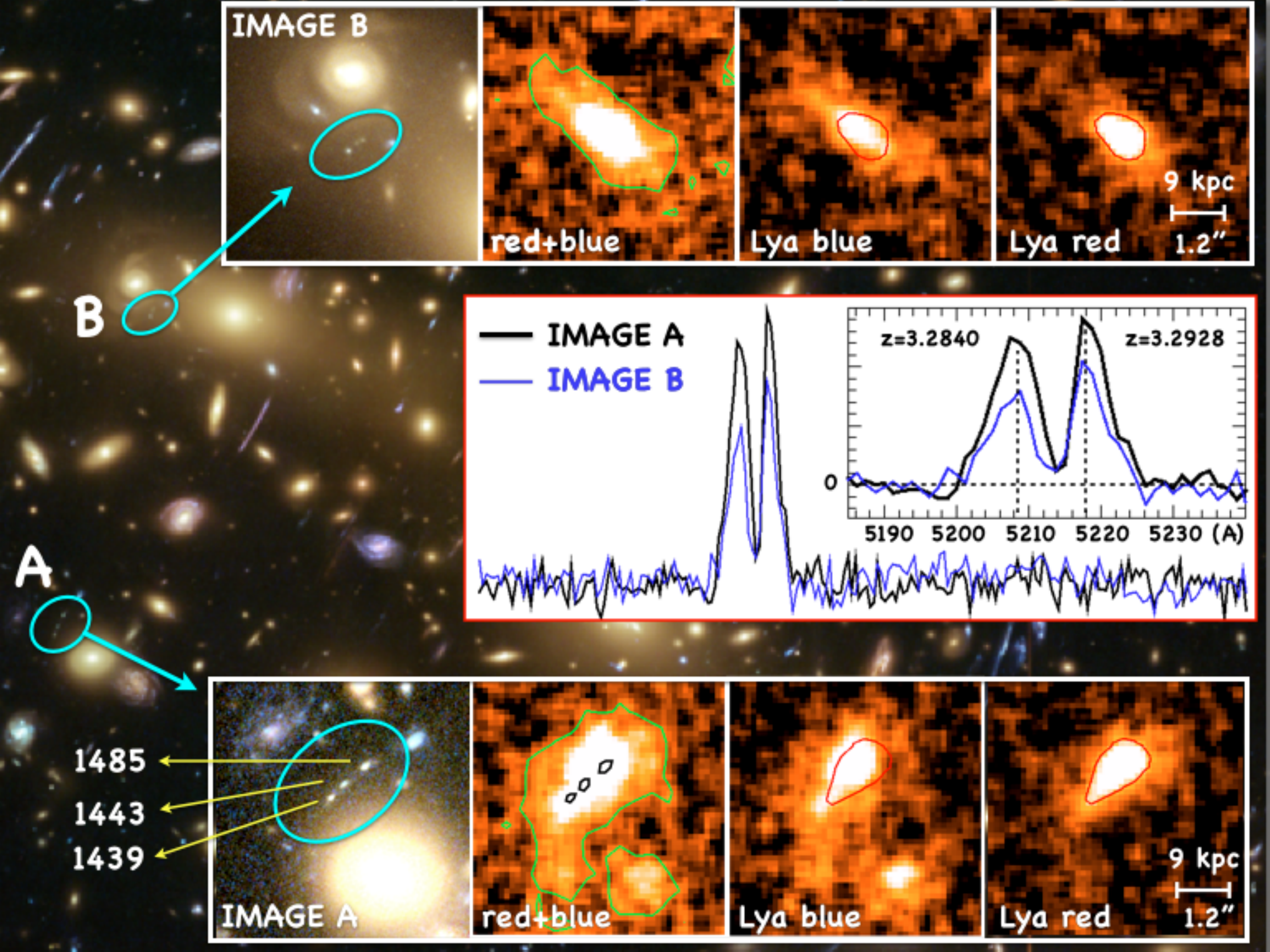} 
\caption{The lensed Ly$\alpha$ nebula discovered behind the Hubble Frontier FIeld MACSJ0416
at $z=3.2$ is shown. Two multiple images have been marked and Ly$\alpha$ spatial distribution is shown
in the insets (from left to right: the color image, Ly$\alpha$ "red + blue", only red and only blue).
In the middle inset the spectrum around the Ly$\alpha$ and a zoomed spectral profile are shown for both
images, A (thick black line) and B (thick red line). The three galaxies associated to the nebula are
marked in the bottom left with yellow arrows and the corresponding IDs from the ASTRODEEP catalog \citep{castellano16}.}
\label{lensed}
\end{figure*}

\begin{figure*}
\centering
\includegraphics[width=16cm]{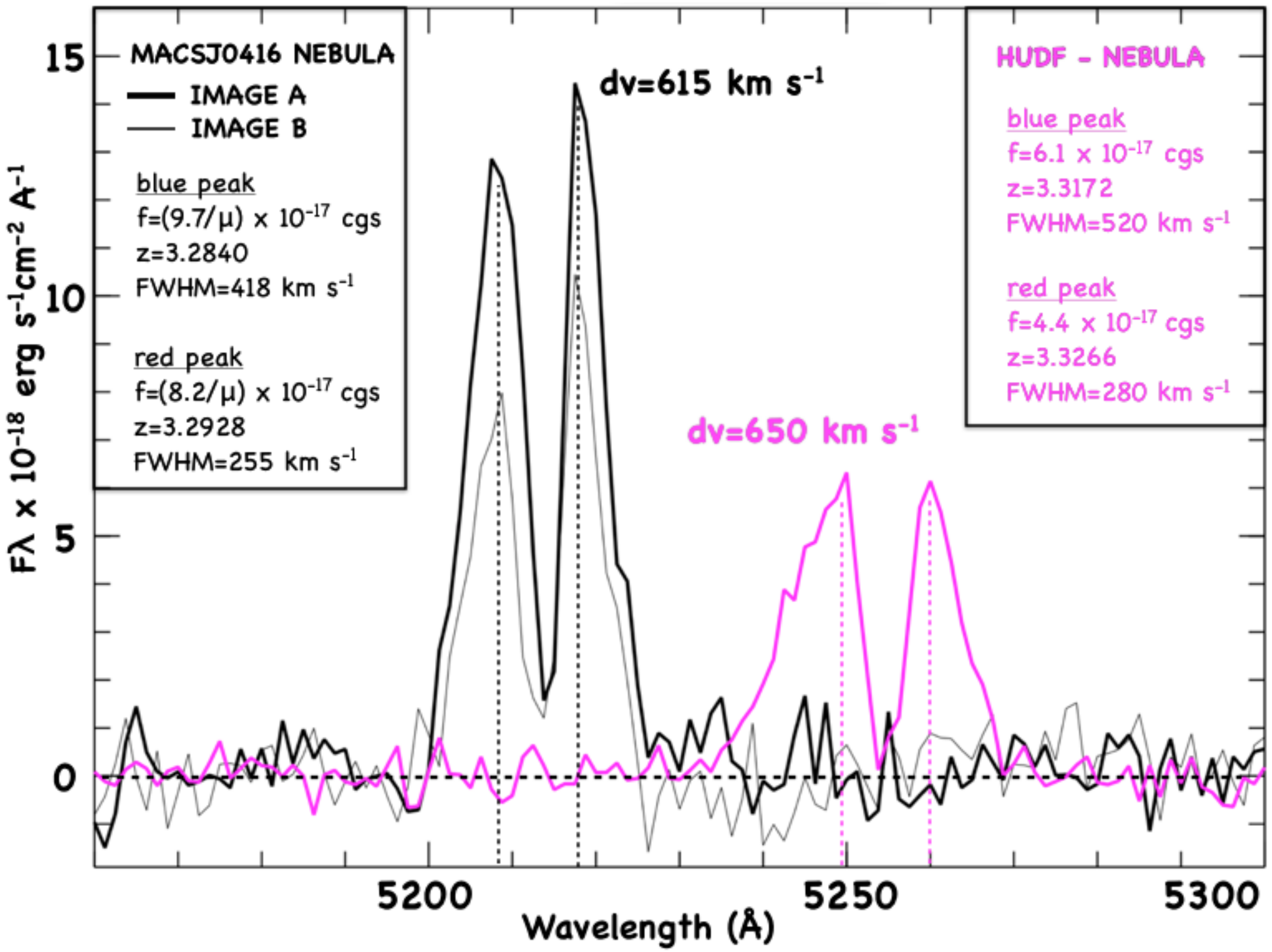} 
\caption{The two Ly$\alpha$ profiles of the nebulae described in this work are shown in the
observed wavelength domain. Black thick (thin) line shows the profile for the lensed nebula, image A (B).
Relevant parameters are reported. In the case of MACSJ0416 nebula numbers are
reported for image A, for which the magnification is $\mu \simeq2$ (Caminha et al. in prep.).
Remarkably, the trough, velocity difference, and widths of the two nebulae are very similar.}
\label{comparison}
\end{figure*}

\subsection{Gravitationally lensed Ly$\alpha$ emitting nebula in the Hubble Frontier Fields MACS~J0416}

Another strongly magnified Ly$\alpha$ nebula has been discovered at redshift $z=3.22$
behind the Hubble Frontier Fields MACSJ0416, as a gravitationally lensed system showing two multiple images
(see Figure~\ref{lensed}).  We report about this system because of its very similar spectral properties to the
nebula discovered in the HUDF.
The size of the emitting gas is smaller
than the system discovered in the HUDF, having a lens-uncorrected (i.e., observed) extension of
the order of 40 kpc. The two images have been identified
at RA=04h16m10.9s, DEC=-24$^{\circ}$04$'$20.7$"$ (the more magnified image, image A hereafter,
with magnification $\mu \simeq 2$) and RA=04h16m09.6s, DEC=-24$^{\circ}$03$'$59.7$"$
(a second less magnified image, image B hereafter, with $\mu \simeq 1$).
This multiple imaged system has been used in the new highly-precise strong lensing model,
which uses a large set of new multiple systems spectroscopically confirmed with MUSE
(it corresponds to System 9 in Caminha et al. in prep.). The image positions are fully consistent with the
strong lensing model, which accurately reproduces their positions.
The morphology in the source plane is slightly distorted, but a more careful dedicated 
modeling would be required to derive meaningful conclusions. That is out of the scope of the present work 
and it will be presented in a separate dedicated study (Caminha et al. in prep.). We note that
also in this case the spatial distribution of Ly$\alpha$-red and Ly$\alpha$-blue (especially in
image A) are not coincident. This is shown in Figure~\ref{lensed}, where the red contours mark the
position of the Ly$\alpha$-red emission, which appears slightly displaced compared to the 
Ly$\alpha$-blue emission).

Remarkably, the spectral properties of this nebula are very similar to those of the HUDF nebula
discussed above. 
In particular, the (1) rest-frame velocity difference, (2) line widths (FWHM), and (3) global shape  
-- e.g., the bluer Ly$\alpha$ peak is systematically broader than the red one --  
are very similar in both nebulae. Figure~\ref{comparison} shows a comparison of the two
spectral profiles as they are observed. After correcting the line fluxes
by the magnification factor (e.g., $\mu = 2$ for image A), we obtain a Ly$\alpha$
luminosity of $4.4 \times 10^{42}$\ergs\,  and $3.7 \times 10^{42}$\ergs\, for the blue and 
red peaks, respectively. The relative peak ratio changes as in the HUDF nebula, showing
different regions where either the blue or the red emission dominate.

The multiple images produced by strong lensing allow us to investigate the broad-band
counterparts more easily than in the HUDF case. In particular a triplet of galaxies
aligned with the emitting gas (Ly$\alpha$) have been identified. They are visible in both multiple images 
(A and B) and, therefore, are likely associated to the nebula.
If they were at different redshift no multiple images would be produced or they would be significantly 
offset from the Ly$\alpha$ emission (see Figure~\ref{lensed}). It is worth noting that also in the strongly
lensed Ly$\alpha$ blob studied by \citet{caminha16} there were three aligned star-forming galaxies
associated to the emitting gas.
The clear double peaked spectral shape of the Ly$\alpha$ emission showing blue and red 
asymmetries suggest it is not the simple sum of the Ly$\alpha$ emissions arising from each
galaxy and blurred by the seeing in the MUSE image. Rather, it appears as a diffuse well resolved emission 
of $\simeq 4.0\times2.5$\arcsec\ (in the case of image A) detected in a region much larger than the stellar
continuum of each galaxy (see black contours in Figure~\ref{lensed}).

Other than the Ly$\alpha$ emission, no additional spectral feature (e.g., no high ionisation lines, 
\civ, \oiiiuv, \ciiidoub\ have been identified in the spectra of these galaxies. 
Therefore, we assume  
that these three galaxies are at the redshift of the nebula on the basis of geometrical arguments,
and with the support of the lensing model.
We checked the photometric redshifts of the three galaxies using the ASTRODEEP
photometric redshift catalog recently published by \citet{castellano16}. All of them
are fully consistent with the redshift of the Ly$\alpha$ nebula, $z=3.2$. More specifically, 
the three galaxies have ASTRODEEP IDs 1439, 1443 and 1485 and photometric
redshift $\simeq 3.55, 3.35$ and 3.45.
Two of them, 1439 and 1443, show relatively low stellar mass $4.4\times10^{8} M_{\odot}$
and $3.0\times10^{9} M_{\odot}$ and a possible excess in the K-band that would suggest 
a strong nebular contribution in this band by the group of lines \oiiidoub\ + H$\beta$.
As discussed above, these features may suggest a possible link with escaping ionizing 
radiation. The third galaxy, 1485, is the most massive one with $1.5\times10^{10} M_{\odot}$.

The star formation rate of the three objects (1439, 1443, and 1485)
has been estimated to be 1.5, 1.2, and 3.4\msunyr, respectively \citep{castellano16}.

\section{Discussion and Conclusions}

We discovered and discussed two extended Ly$\alpha$ systems at redshift $\simeq$ 3.3. The prominent blue peak in their Ly$\alpha$ spectra
accompanied by a fainter red, slightly narrower peak is remarkable. Usually, Ly$\alpha$ has been observed with dominant red tails indicative of
outflows \citep[e.g.,][]{shapley03,vanzella09} as well as many Ly$\alpha$ blobs \citep[e.g.,][]{matsuda06}. Here we observe the opposite.

This is even more relevant if the intergalactic absorption is considered,  that would tend to preferentially suppress the bluer peak.
In particular, an IGM transmission in the blue side of the Ly$\alpha$ ranging between 20\% and 95\% (68\% interval) with a mean of
$\sim 80$\% has been proposed by \citet{laursen11}.

The Ly$\alpha$ nebulae described in this work benefit of the integral field spectroscopy (MUSE), which is more informative that previous
long-slit studies \citep[e.g.,][]{rauch16}. Despite that, the complexity of the system still prevent us from deriving firm conclusions. So we can at
best test the plausibility of the processes involved and discuss the most likely scenarios.

As mentioned above the nebulae described in this work show quite complex
structure with spatial-depended Ly$\alpha$ emission (Figures~\ref{lyablue} and \ref{lyared}) 
and varying sub-spectral profiles (Figure~\ref{lyaprofile}), 
however two clear spectral features are present in both systems:
(1) the broad doubled peaked line profile with prominent blue emission and 
(2) the `trough' separating the two peaks which appears to occur mostly at the
same frequency throughout the nebula (this is best illustrated by the {\it left panel} of Figure~\ref{lyaprofile}).
These two observations combined can be naturally explained with scattering, and
support the fact that radiative transfer effects are likely responsible for shaping the spectra emerging from these nebulae. 

The simplest version of the scattering medium consists of a static gas cloud of uniform density. For such a medium, the emerging
Ly$\alpha$ spectrum is double peaked, with the peak separation set by the total HI column density of the cloud 
\citep[e.g.,][]{harrington73,neufeld90,dijkstra14}:

\begin{equation}
N_{HI} =5.3\times10^{20} \left( \frac{T}{10^{4}K} \right) ^{-1/2} \left( \frac{{\rm dv}}{670\hspace{1mm} {\rm km}\hspace{1mm} {\rm s}^{-1}} \right) ^{3} {\rm cm}^{-2}. 
\end{equation} 

However, this possibility has been excluded by \citet{rauch11} on the basis of similar spectral properties we find here, e.g.,
the relatively stronger intensity of the blue versus red emission and the different widths suggests that we are not observing a static
configuration. Radiative transfer
through clumpy/multiphase media can also explain double peaked spectra \citep[see ][]{gronke16}. Clumpy media generally give rise to
a wide variety of broad, multi-peaked spectral line profiles. The trough at a constant frequency then reflects either that there is a non-negligible opacity
in residual HI in the hot inter clump gas  \citep{gronke16}, or in the cold clumps that reside in the hot halo gas (the presence of these clumps
inside massive dark matter halos has been proposed by, e.g., \citealt{cantalupo14} and \citealt{hennawi15}). The width of the though reflects
either the thermal broadening of the Ly$\alpha$ absorption cross-section for residual HI in the hot gas, and/or the velocity dispersion of the cold clumps.\\

Also the origin of Ly$\alpha$ emission in these nebulae is not easily identifiable. 

{\bf Star Formation.} The association of the brightest Ly$\alpha$ spots with galaxies (in the HUDF nebula) and the three aligned star-forming
galaxies in the lensed nebula suggest that star formation is at least powering the Ly$\alpha$ emission from the high-surface brightness spots. 
The enhancement of the blue and/or red peak in these brighter spots either reflects the kinematics of the clumps surrounding these galaxies,
where an enhanced blue (red) peak is indicative of clumps falling onto (flowing away from) the galaxies 
\citep[see][]{zheng02,dijkstra06a,dijkstra06b,verhamme06}. Alternatively, enhancements in any of the two peaks could be due the 
galaxies bulk motion with respect to the Ly$\alpha$ emitting nebula (i.e. these could be star-forming galaxies which are moving onto the
more massive halo that hosts the Ly$\alpha$ nebula).\\

{\bf Cooling.} For massive dark matter halos ($M_{\rm halo}\sim 1-5 \times 10^{12}$M$_{\odot}$), the cooling luminosity can reach $L \sim 10^{43}$ erg s$^{-1}$ 
\citep[e.g.,][]{dijkstra09,giguere10,goerdt10,rosdahl12}, which is sufficient to explain these halos. The overall double
peaked spectrum is reminiscent of that predicted by \citet{giguere10} for cooling radiation (though see \citealt{trebitsch16}). 
In particular the observed velocity difference ($\simeq 650$\kms) and the Ly$\alpha$ total luminosity are compatible to those reported by \citet{giguere10}. 
Also in terms of spatial emission there are similarities, like the extensions of several tens kpc and the emission arising in different places. The spatial extent results from
a combination of some of the cooling radiation being emitted in the accretion streams far from the galaxy(ies), and of spatial diffusion owing to resonant scattering.
Also the presence of a well developed blue component slightly broader than the red one is among the outputs of the \citet{giguere10} prescriptions 
(see their Figure~5), and is indicative of systematic infall, in which the velocity of the in-flowing gas tend to move (in the frequency domain) the red Ly$\alpha$ photons
toward the resonance, while making easier for the blue Ly$\alpha$ photons to escape in the blue side. \\

{\bf Fluoresence.} It is also possible that ionising photons escape from galaxies, which would cause the cold clouds to fluoresce in Ly$\alpha$
\citep[e.g.,][]{mas16}. Fluorescent Ly$\alpha$ emission also gives rise to a double peaked spectrum, in which infalling/outflowing material diminishes the red/blue wing of the spectrum. However, fluorescence tend to produce a smaller peak separation
than what is observed here \citep[e.g.,][]{gould96,cantalupo05}. In other words, if fluorescence is the source of Ly$\alpha$ emission, 
then we would still need additional scattering to explain the width of the Ly$\alpha$ line. 
Note that this explanation also relies on star formation powering the Ly$\alpha$ emission. Given the number of galaxies that are possibly
associated to each system, this explanation is energetically compatible with the observed star formation activity \citep[see][]{rauch11}. 
Moreover, as mentioned above, the strong optical rest-frame nebular emission (\oiiidoub\ and H$\beta$) as traced by the K-band for some of the
galaxies (e.g. \#3, \#2, and \#6 in the HUDF nebula and  in the lensed nebula) is intriguingly similar  to what has been recently observed in a $z=3.2$
galaxy (with equivalent widths of EW(\oiiidoub)=1500\AA, \citealt{debarros16} and showing a remarkable
amount of escaping ionizing radiation (higher than 50\%, \citealt{vanzella16}). Among them it is worth noting the presence of the extremely blue (and in practice dust-free), 
compact, and low-mass galaxy (\#6), that might further contribute to the ionisation budget.
The Ly$\alpha$ resonance emission from nebulae may be indirect probes of the escaping ionising radiation from the embedded sources (even fainter that the detection limit),
along transverse directions not accessible from the observer. While, the direct detection of Lyman continuum emission is in principle possible, it might be precluded in the
present data because galaxies are surrounded by the same (circum-galactic) medium plausibly producing the Ly$\alpha$ nebula and therefore preventing us to to easily
detect ionising flux. In addition,  the intergalactic opacity in the Lyman continuum also affects these measurements \citep[e.g.,][]{vanzella12},  requiring a larger sample
of similar systems to average the IGM stochastic attenuation.

{\bf AGN Activity.} AGN can inject large amounts of energy in the surrounding gas \citep[e.g.,][]{debuhr12}, which could radiate even after the nucleus has
shut off. Such processes can modify the kinematic and thermal properties of the circum-galactic medium, and therefore its Ly$\alpha$ signatures.

The key observable that distinguishes between different sources of Ly$\alpha$ emission would be the Balmer emission lines (like H$\alpha$ or H$\beta$). In the case of
cooling radiation, the H$\alpha$ flux that is associated with Ly$\alpha$ should be $\sim $ 100 times weaker \citep{dijkstra14}, and would likely be undetectable. However,
for the `star formation' and `fluorescence' models, the H$\alpha$ flux should be significantly stronger. If H$\alpha$ emission can be observed, and confined to galaxies
then Ly$\alpha$ emission was likely powered by nebular emission inside galaxies, while fluorescence would give rise to partially extended H$\alpha$.

The search for galaxies that illuminate themselves through some fortuitous release of Ly$\alpha$
or ionizing radiation into their environment offer positive prospects in the future MUSE observations 
and will provide our main direct insights into the in- and outflows of gas \citep{rauch16}. 
If the scenario in which the gas is inflowing toward a region forming stars is correct, then we may be
witnessing an early phase of galaxy or a proto-cluster (or group) formation.
Searches for asymmetric Ly$\alpha$ halos or offsets between stellar populations
and Ly$\alpha$ emission may reveal further objects where the
escape of ionizing radiation can be studied.

\bigskip 

\section*{Acknowledgements}

We thanks F. Calura and G. Zamorani for useful discussions and A. Grazian for providing us the K-band
image of the HUDF nebula. Part of this work has been funded through the PRIN INAF 2012.

\bsp	
\label{lastpage}

\begin{thebibliography}{}
  
\bibitem[Adelberger et al.(2003)]{adelberger03} Adelberger, K.~L., Steidel, C.~C., Shapley, A.~E., \& Pettini, M.\ 2003, \apj, 584, 45
\bibitem[Bacon et al.(2012)]{bacon12} Bacon, R., Accardo, M., Adjali, L., et al.\ 2012, The Messenger, 147, 4 
\bibitem[Beckwith et al.(2006)]{beckwith06} Beckwith, S.~V.~W., Stiavelli, M., Koekemoer, A.~M., et al.\ 2006, \aj, 132, 1729 
\bibitem[Borisova et al.(2016)]{borisova16} Borisova, E., Cantalupo, S., Lilly, S.~J., et al.\ 2016, arXiv:1605.01422 
\bibitem[Caminha et al.(2015)]{caminha16} Caminha, G.~B., Karman, W., Rosati, P., et al.\ 2015, arXiv:1512.05655
\bibitem[Cantalupo et al.(2005)]{cantalupo05} Cantalupo, S., Porciani, C., Lilly, S.~J., \& Miniati, F.\ 2005, \apj, 628, 61 
\bibitem[Cantalupo et al.(2014)]{cantalupo14} Cantalupo, S., Arrigoni-Battaia, F., Prochaska, J.~X., Hennawi, J.~F., \& Madau, P.\ 2014, \nat, 506, 63
\bibitem[Castellano et al.(2016)]{castellano16} Castellano, M., Amor{\'{\i}}n, R., Merlin, E., et al.\ 2016, \aap, 590, A31 
\bibitem[Chen et al.(2001)]{chen01} Chen, H.-W., Lanzetta, K.~M., Webb, J.~K., \& Barcons, X.\ 2001, \apj, 559, 654 
\bibitem[Christensen et al.(2004)]{christensen04} Christensen, L., S{\'a}nchez, S.~F., Jahnke, K., et al.\ 2004, \aap, 417, 487 
\bibitem[Coe et al.(2006)]{coe06} Coe, D., Ben{\'{\i}}tez, N., S{\'a}nchez, S.~F., et al.\ 2006, \aj, 132, 926 
\bibitem[de Barros et al.(2016)]{debarros16} de Barros, S., Vanzella, E., Amor{\'{\i}}n, R., et al.\ 2016, \aap, 585, A51 
\bibitem[Debuhr et al.(2012)]{debuhr12} Debuhr, J., Quataert, E., \& Ma, C.-P.\ 2012, \mnras, 420, 2221 
\bibitem[Dijkstra et al.(2006b)]{dijkstra06b} Dijkstra, M., Haiman, Z., \& Spaans, M.\ 2006, \apj, 649, 37 
\bibitem[Dijkstra et al.(2006a)]{dijkstra06a} Dijkstra, M., Haiman, Z., \& Spaans, M.\ 2006, \apj, 649, 14 
\bibitem[Dijkstra \& Loeb(2009)]{dijkstra09} Dijkstra, M., \& Loeb, A.\ 2009, \mnras, 400, 1109 
\bibitem[Dijkstra(2014)]{dijkstra14} Dijkstra, M.\ 2014, \pasa, 31, e040
\bibitem[Faucher-Gigu{\`e}re et al.(2010)]{giguere10} Faucher-Gigu{\`e}re, C.-A., Kere{\v s}, D., Dijkstra, M., Hernquist, L., \& Zaldarriaga, M.\ 2010, \apj, 725, 633 
\bibitem[Fontana et al.(2014)]{fontana14} Fontana, A., Dunlop, J.~S., Paris, D., et al.\ 2014, \aap, 570, A11 
\bibitem[Francis et al.(2013)]{francis13} Francis, P.~J., Dopita, M.~A., Colbert, J.~W., et al.\ 2013, \mnras, 428, 28 
\bibitem[{{Giavalisco} {et~al.}(2004){Giavalisco}, {Ferguson}, {Koekemoer},
  {Dickinson}, {Alexander}, {Bauer}, {Bergeron}, {Biagetti}, {Brandt},
  {Casertano}, {Cesarsky}, {Chatzichristou}, {Conselice}, {Cristiani}, {Da
  Costa}, {Dahlen}, {de Mello}, {Eisenhardt}, {Erben}, {Fall}, {Fassnacht},
  {Fosbury}, {Fruchter}, {Gardner}, {Grogin}, {Hook}, {Hornschemeier}, {Idzi},
  {Jogee}, {Kretchmer}, {Laidler}, {Lee}, {Livio}, {Lucas}, {Madau},
  {Mobasher}, {Moustakas}, {Nonino}, {Padovani}, {Papovich}, {Park},
  {Ravindranath}, {Renzini}, {Richardson}, {Riess}, {Rosati}, {Schirmer},
  {Schreier}, {Somerville}, {Spinrad}, {Stern}, {Stiavelli}, {Strolger},
  {Urry}, {Vandame}, {Williams}, \& {Wolf}}]{giava04}
\bibitem[Giavalisco et al.(2011)]{giavalisco11} Giavalisco, M., Vanzella, E., Salimbeni, S., et al.\ 2011, \apj, 743, 95 
\bibitem[Goerdt et al.(2010)]{goerdt10} Goerdt, T., Dekel, A., Sternberg, A., et al.\ 2010, \mnras, 407, 613 
\bibitem[Gould \& Weinberg(1996)]{gould96} Gould, A., \& Weinberg, D.~H.\ 1996, \apj, 468, 462 
\bibitem[Gronke \& Dijkstra(2016)]{gronke16} Gronke, M., \& Dijkstra, M.\ 2016, arXiv:1604.06805 
\bibitem[Guo et al.(2013)]{guo13} Guo, Y., Ferguson, H.~C., Giavalisco, M., et al.\ 2013, \apjs, 207, 24 
\bibitem[Hayes et al.(2011)]{hayes11} Hayes, M., Scarlata, C., \& Siana, B.\ 2011, \nat, 476, 304 
\bibitem[Hennawi et al.(2015)]{hennawi15} Hennawi, J.~F., Prochaska, J.~X., Cantalupo, S., \& Arrigoni-Battaia, F.\ 2015, Science, 348, 779 
\bibitem[Harrington(1973)]{harrington73} Harrington, J.~P.\ 1973, \mnras, 162, 43 
\bibitem[Lanzetta et al.(1995)]{lanzetta95} Lanzetta, K.~M., Bowen, D.~V., Tytler, D., \& Webb, J.~K.\ 1995, \apj, 442, 538
\bibitem[Laursen et al.(2011)]{laursen11} Laursen, P., Sommer-Larsen, J., \& Razoumov, A.~O.\ 2011, \apj, 728, 52 
\bibitem[Lotz et al.(2014)]{lotz14} Lotz, J., Mountain, M., 
Grogin, N.~A., et al.\ 2014, American Astronomical Society Meeting Abstracts \#223, 223, 254.01 
\bibitem[Karman et al.(2016)]{karman16} Karman, W., Caputi, K.~I., Caminha, G.~B., et al.\ 2016, arXiv:1606.01471 
\bibitem[Koekemoer et al.(2014)]{kokom14} Koekemoer, A.~M., Avila, R.~J., Hammer, D., et al.\ 2014, American Astronomical Society 
Meeting Abstracts \#223, 223, 254.02 
\bibitem[Kulas et al.(2012)]{kulas12} Kulas, K.~R., Shapley, A.~E., Kollmeier, J.~A., et al.\ 2012, \apj, 745, 33
\bibitem[Martin et al.(2015)]{martin15} Martin, D.~C., Matuszewski, M., Morrissey, P., et al.\ 2015, \nat, 524, 192 
\bibitem[Mas-Ribas \& Dijkstra(2016)]{mas16} Mas-Ribas, L., \& Dijkstra, M.\ 2016, \apj, 822, 84 
\bibitem[Matsuda et al.(2006)]{matsuda06} Matsuda, Y., Yamada, T., Hayashino, T., Yamauchi, R., \& Nakamura, Y.\ 2006, \apjl, 640, L123 
\bibitem[Neufeld(1990)]{neufeld90} Neufeld, D.~A.\ 1990, \apj, 350, 216 
\bibitem[Nonino et al.(2009)]{nonino09} Nonino, M., Dickinson, M., Rosati, P., et al.\ 2009, \apjs, 183, 244 
\bibitem[Patr{\'{\i}}cio et al.(2016)]{patricio16} Patr{\'{\i}}cio, V., Richard, J., Verhamme, A., et al.\ 2016, \mnras, 456, 4191 
\bibitem[Prescott et al.(2012)]{prescott12} Prescott, M.~K.~M., Dey, A., Brodwin, M., et al.\ 2012, \apj, 752, 86 
\bibitem[Prescott et al.(2015)]{prescott15} Prescott, M.~K.~M., Martin, C.~L., \& Dey, A.\ 2015, \apj, 799, 62 
\bibitem[Rafelski et al.(2015)]{rafelski15} Rafelski, M., Teplitz, H.~I., Gardner, J.~P., et al.\ 2015, \aj, 150, 31 
\bibitem[Rauch et al.(2011)]{rauch11} Rauch, M., Becker, G.~D., Haehnelt, M.~G., et al.\ 2011, \mnras, 418, 1115 
\bibitem[Rauch et al.(2016)]{rauch16} Rauch, M., Becker, G.~D., \& Haehnelt, M.~G.\ 2016, \mnras, 455, 3991 
\bibitem[Rosdahl \& Blaizot(2012)]{rosdahl12} Rosdahl, J., \& Blaizot, J.\ 2012, \mnras, 423, 344 
\bibitem[Shapley et al.(2003)]{shapley03} Shapley, A.~E.,  Steidel, C.~C., Pettini, M., \& Adelberger, K.~L.\ 2003, \apj, 588, 65
\bibitem[Steidel et al.(2010)]{steidel10} Steidel, C.~C., Erb, D.~K., Shapley, A.~E., et al.\ 2010, \apj, 717, 289 
\bibitem[Swinbank et al.(2015)]{swinbank15} Swinbank, A.~M., Vernet, J.~D.~R., Smail, I., et al.\ 2015, \mnras, 449, 1298 
\bibitem[Trebitsch et al.(2016)]{trebitsch16} Trebitsch, M., Verhamme, A., Blaizot, J., \& Rosdahl, J.\ 2016, arXiv:1604.02066 
\bibitem[Turner et al.(2014)]{turner14} Turner, M.~L., Schaye, J., Steidel, C.~C., Rudie, G.~C., \& Strom, A.~L.\ 2014, \mnras, 445, 794 



\bibitem[Vanzella et al.(2009)]{vanzella09} Vanzella, E., Giavalisco, M., Dickinson, M., et al.\ 2009, \apj, 695, 1163 

\bibitem[{{Vanzella} {et~al.}(2012){Vanzella}, {Guo}, {Giavalisco}, {Grazian},
  {Castellano}, {Cristiani}, {Dickinson}, {Fontana}, {Nonino}, {Giallongo},
  {Pentericci}, {Galametz}, {Faber}, {Ferguson}, {Grogin}, {Koekemoer},
  {Newman}, \& {Siana}}]{vanzella12}
{Vanzella}, E., {Guo}, Y., {Giavalisco}, M., {et~al.} 2012, \apj, 751, 70, 70

\bibitem[Vanzella et al.(2016)]{vanzella16} Vanzella, E., de Barros, S., Vasei, K., et al.\ 2016, \apj, 825, 41 
\bibitem[Verhamme et al.(2006)]{verhamme06} Verhamme, A., Schaerer, D., \& Maselli, A.\ 2006, \aap, 460, 397 


\bibitem[Wisotzki et al.(2016)]{wisotzki16} Wisotzki, L., Bacon, R., Blaizot, J., et al.\ 2016, \aap, 587, A98 
\bibitem[Zheng \& Miralda-Escud{\'e}(2002)]{zheng02} Zheng, Z., \& Miralda-Escud{\'e}, J.\ 2002, \apj, 578, 33 

\end{thebibliography}
\end{document}